\documentclass[10pt]{article}
\usepackage[latin1]{inputenc}
\usepackage{amsmath}
\usepackage{amsfonts}
\usepackage{amssymb}
\usepackage{mathrsfs}
\usepackage{subfig}
\usepackage{fancybox,graphicx}
\usepackage{subfig}
\usepackage{caption}
\usepackage{color}
\usepackage{authblk}
\usepackage[colorlinks]{hyperref}
\usepackage{accents}
\usepackage[titletoc,title]{appendix}
\usepackage{cite}

\usepackage[top=2in, bottom=1.5in, left=1in, right=1in]{geometry}

\newcommand{\pvec}[1]{\vec{#1}\mkern2mu\vphantom{#1}} 

\title{Analytic treatment of vortex motion in non-inertial Stokes flows} 

\author[1,2]{Robert Salazar}
\author[2,3]{Camilo Bayona}
\affil[1]{Departamento de F\'isica, Universidad de los Andes - Bogot\'a, Colombia}
\affil[2]{Universidad ECCI - Bogot\'a, Colombia}
\affil[3]{Centro de Ingenier\'ia Avanzada Investigaci\'on y Desarrollo, CIAID- Bogot\'a, Colombia}

\begin{document}
    \maketitle

\begin{abstract}
We study the steady state motion of incompressible and viscous fluid flow in a rotating reference frame where vortices may take place. An approximated analytic solution of the Stokes flow problem is proposed for situations where the vorticity is highly concentrated along a given direction. The approximation disconnects the component of velocity along the axis of rotation from the momentum equation. This enables to find the vorticity from the solution of the Poisson's equation. We use the Green's function approach to provide solutions of the three-dimensional flow on cylindrical domains with radial inlet/outlet velocity profiles by using an exact expansion of the Green's function. This approximated analytical solution for the vorticity is in good agreement with numerical solutions computed with Finite Difference Method (FDM).               
\end{abstract}

\section{Introduction}
The study  of three-dimensional vortex motion plays an important role in the hydrodynamic
description of diverse systems ranging from the 3D driven cavity to tropical cyclones in the
earth's atmosphere. The analytic description of the vortex motion is a challenging problem
because precisely, it requires to solve the Navier-Stokes equations. Nowadays, such
governing equations of the hydrodynamics do not have an universal solution, and remain
as an open problem. The main difficulty concerns the treatment of the mathematical problem for high values of the number of the Reynolds number $Re$ where the non-linear terms of the governing equations become important. However, there are particular solutions of the Navier-Stokes equations \cite{wang1989exact,wang1991exact,okamoto1997exact,nugroho2010class}, among them a well-known solution is the Burger's vortex \cite{BurgersVortex1948} which gives a stationary exact description of the three-dimensional problem. Sullivan (1959) obtained an exact solution to the Navier-Stokes equation in the form of a steady two-cell vortex \cite{roger1959two}. Bellamy-Knights (1970) generalized the Sullivan solution for a time-dependent situation. The Sullivan vortex exactly coincides with the Burgers vortex far from the vortex's core. The analytic vortex models are also important because they may be used to study complex phenomena as turbulence \cite{ashurst1987alignment,andreotti1997studying} as well as models of the tangent velocity profile of cyclones in the atmosphere \cite{wood2011new} or in daily situations as the bath tube vortex. Although, analytic solutions are valid in particular situations, such solutions have important implications in the field of Computational Fluid Dynamics (CFD), namely the Finite Element Method (FEM) since analytic solutions are required not only to test the accuracy of the numerical approximations, but also as theoretical knowledge for constructing closure models for turbulence modelling \cite{andreotti1997studying}.

A linear version of the Navier-Stokes equations is the Stokes flow problem. This type of flow occurs for low values of $Re$ where the advective terms of the Navier-Stokes equations can be vanished and the system, in principle, never reaches the turbulent regime. The governing equations for Stokes flows do not a have general analytical solution even when they are simpler than the Navier-Stokes equations. This type of flow take place in viscous fluids at low velocities and there are several analytical studies about Stokes flows description in diverse situations   \cite{gaskell1998stokes,masoud2009analytical,wang2014symplectic,herschlag2015exact,
okabe1983analytical,burda2012analytical,crowdy2017analytical,crowdy2017analytical,
crowdy2010stokes,shankar2009exact,shankar2009exact,zhong1996analytic}.

In this document, we study the motion of a highly viscous and incompressible rotating fluid. This problem can be modeled as a stationary Stokes flow occurring on a non-inertial reference frame. The main goal of the current work is to find approximated analytic solutions of the linearized Navier-Stokes equations. This problem has practical interest since an important source of vorticity on these rotating fluids comes from the Coriolis force as occurs in some geophysical systems e.g. \textit{cyclones} in the atmosphere or \textit{gyres} in the oceans. We show that steady vortices may take place as a consequence of the continuity and the non-inertial forces when motion of fluid along the rotation axis is provided. We propose an approximation valid in the situation where a component of vorticity is mainly concentrated along the axis defined by the angular velocity of the rotating frame. We find integral analytic expressions for the vorticity and the velocity field whose evaluation can be performed by numerical integration, especially in the axially symmetric case. In practice, the approximation simplifies the mathematical problem since it implies that the $z$-velocity $v_z$ (component along the reference frame's axis) must satisfy the Laplace's equation. Under this approximation the $z$-vorticity can be obtained by solving a Poisson's equation depending on $\Pi = \partial_z v_z$. Therefore, if we know a $z$-velocity stationary profile which is a solution of the Laplace's equation, then we may use the approximation to find the other two components of the stationary velocity.   

Here we shall show that a column of ascending fluid may generate a vortex with a variable vorticity along the $z$-axis. The structure of the vortex depends strongly on $\Pi$ since the vortex's spin is determined by the sign of this function. This feature is interesting because it is closely related with the cyclonic/anti-cyclonic directions of hurricane rotation wind in the lower and upper parts of this system. The problem of cyclonic/anti-cyclonic changing directions of a hurricane was modeled analytically in Ref.~\cite{leonov2014analytical} where the diffusive term of the momentum equations can be neglected. In this article, we provide an analytical explanation of the vorticity sign inversion along the axis of rotation for a particular tridimensional vortex in a highly viscous fluid.              

The document is organized as follows. The governing equations for incompressible rotation fluids are presented in Section~\ref{Equations_of_motionSection}. Since we deal with a situation where Coriolis force is the main source of vorticity, then we propose in Section~\ref{Highly_condensed_vorticity_approximationSection} an approximation valid when vorticity is mainly pointing along the axis of rotation of the non-inertial reference frame, let us say the $z$-axis. The Laplace's equation for $v_z$ is solved in Section \ref{The_stationary_z-velocitySection} with Gaussian inlet/outlet velocity profiles in order to use it in the methodology description of the next sections. Section \ref{Integral_form_of_the vorticitySection} is devoted to the determination of an integral form of the vorticity by using the Green's function method. We use a compact expansion of the Green's function in terms of the elliptic integral of the first kind for axially symmetric flow to simplify the problem in cylindrical domains. Next section is focused on the computation of vorticity by using a discrete approximation of $\Pi$ in the space. In particular, $z$-vorticity is computed by using the $z$-velocity profile found in \ref{The_stationary_z-velocitySection}. The radial an tangent velocities are computed in Subsections 
\ref{Radial_velocitySubsection} and \ref{Angular_velocitySubSection} assuming Gaussian inlet/outlet $z$-velocity profiles. Finally, we solve the Poisson's equation for the vorticity in Section \ref{Numerical_solution_of_the_Poisson_EquationEqSection} by using the \textit{Finite Difference Method} (FDM) to perform a comparison between numerical results and the analytic approximations obtained in Sections \ref{Integral_form_of_the vorticitySection} and \ref{Point-wise_distribution_approximation_for_Pi_functionSection}.    

\section{Equations of motion}
\label{Equations_of_motionSection}
The equations of motion are \cite{kundu1990fluid}
\begin{equation}
\partial_t v_\alpha + v_{\beta} \partial_\beta v_{\alpha} = - g \delta_{\alpha,3} + 2\Omega e_{\alpha\beta 3}v_{\beta} + \Omega^2 r_{\bot\alpha} - \frac{1}{\rho} \partial_\alpha P + \nu \partial_\beta \partial_\beta v_{\alpha}
\label{navierStokesEq}
\end{equation}
in the above, the summation convention has been applied where $\alpha,\beta = 1, 2, 3$ corresponds to the spatial dimensions in Cartesian coordinates $x$, $y$ and $z$ respectively, $v_\alpha$ is the velocity, $P$ is the pressure, $\Omega$ the angular velocity of the non-inertial frame, $g$ the gravity, $\nu$ is the kinematic viscosity,  $\partial_t$ is the eurelian derivative (short-hand notation), $e_{\alpha\beta 3}$ is the Levi-Civita permutation symbol and $\delta_{\alpha,\beta}$ is the Kronecker delta. We are using the $f$-plane approximation where $\Omega$ is pointing to the $z$-axis. We deal with a \textit{Stokes flow problem}, a highly viscous and slowly moving flow where $v_{\beta} \partial_\beta v_{\alpha}$ may be neglected. Since the fluid is incompressible then the continuity equation can be written for convenience as follows 
\[
\partial_\alpha v_{\alpha} = 0  \Rightarrow  \partial_i v_i = \Pi  
\]
with $\Pi := -\partial_3 v_3$ and $i=1,2$. The vorticity equation is obtained by applying the rotational to Eq.~(\ref{navierStokesEq}), this is
\begin{equation}
\partial_t \omega_{\alpha} = 2\Omega \partial_z v_{\alpha} + \nu \partial_\beta \partial_\beta \omega_{\alpha}
\label{vorticityEq}
\end{equation}
where $\omega_\alpha=e_{\alpha \beta \gamma} \partial_\beta v_{\gamma}$ is the vorticity. 
 
\section{Highly condensed vorticity approximation}
\label{Highly_condensed_vorticity_approximationSection}
\subsection{The linear problem} 
Since we have neglected the non-linear convective term, then a linearized version of Eq.~(\ref{navierStokesEq}) is 
\begin{equation}
\partial_t v_\alpha = - g \delta_{\alpha,3} + 2\Omega e_{\alpha\beta 3}v_{\beta} + \Omega^2 r_{\bot\alpha} - \frac{1}{\rho} \partial_\alpha P - \nu e_{\beta\gamma\alpha}\partial_\beta \omega_{\gamma}
\label{linearNavierStokesEq}
\end{equation}
where we have used the property 
\[
\partial_\beta\partial_\beta v_\alpha = \partial_\beta (\partial_\alpha v_\alpha) - e_{\beta\gamma\alpha}\partial_{\beta}\left(  e_{\tilde{\alpha}\tilde{\beta}\gamma}\partial_{\tilde{\alpha}} v_{\tilde{\beta}} \right)
\] 
and the continuity equation. If we suppose that the vorticity is highly concentrated along the axis of rotation, then 
\[
(\vec{\nabla}\times\vec{\omega})_{\alpha} = e_{\beta\gamma\alpha} \partial_\beta \omega_{\gamma} \approx e_{\beta\gamma\alpha} \partial_\beta \omega_{3} 
\]
in cylindrical coordinates the approximation reads
\[
(\vec{\nabla}\times\vec{\omega})_{r} = \frac{1}{r}\partial_\phi \omega_z \hspace{0.5cm}\mbox{,}\hspace{0.5cm}(\vec{\nabla}\times\vec{\omega})_{\phi} = - \partial_r \omega_z \hspace{0.5cm}\mbox{and}\hspace{0.5cm}(\vec{\nabla}\times\vec{\omega})_{z} = 0.
\] 
In principle, the $\phi$ dependence on the vorticity according to equation Eq.~(\ref{vorticityEq}) comes from the $\Pi$. If such function is axially symmetric $\Pi=\Pi(r,z)$, then $\frac{\partial \omega_z}{\partial \phi}=0$. As a result, Eq.~(\ref{linearNavierStokesEq}) will take the form
\[
\partial_t v_r = \Omega^2 r + 2 \Omega v_{\phi} - \frac{1}{\rho} \partial_r P 
\]
\[
\partial_t v_\phi = - 2 \Omega v_{r} - \frac{1}{\rho r} \partial_\phi P + \nu \partial_r \omega_z 
\]
\[
\partial_t v_z = - g + \frac{1}{\rho} \partial_z P    
\]
If we apply the divergence over Eq.~(\ref{linearNavierStokesEq}), we obtain
\[
\vec{\nabla}^2 P = 2\rho\Omega\left[\omega_z - \Omega\frac{\cos\phi+\sin\phi}{2}\right]. 
\]
The Poisson equation for pressure when  $\omega_z>>1$ can be written as follows
\begin{equation}
  \vec{\nabla}^2 P \underset{\omega_z >> 1}{=}  2\rho\Omega \omega_z .  
  \label{PoissonForPWhenOmegaLargeEq}
\end{equation}
Since the $z$-vorticity is axially symmetric, then pressure will be approximately independent of $\phi$ 
specially in such regions where the $z$-vorticity is larger than the term $|(\Omega/2)(\cos\phi+\sin\phi)|$ coming from the centrifugal force.  

\subsection{Steady flow}
If $\omega_z$ is large, the pressure is practically axially symmetric and the steady flow is described by the following equations 
\begin{equation}
v_{\phi} = \frac{1}{2\Omega \rho} \partial_r P - \frac{1}{2} \Omega r 
\label{A-Eq}
\end{equation}
\begin{equation}
v_{r} = - \frac{\nu}{2\Omega} \partial_r \omega_z 
\label{B-Eq}
\end{equation}
\begin{equation}
\partial_z P = -\rho g
\label{C-Eq}
\end{equation}
The last equation represents a hydrostatic pressure balance along the $z$-axis under the assumption that vorticity is highly concentrated on this axis. This type of hydrostatic balance also occurs in the ocean and atmospheric context where the estimation of the order of magnitude lead to the conclusion that vertical pressure gradient is the dominant term of the momentum equation along the $z$-axis \cite{pedlosky2013geophysical,vallis2017atmospheric,bouchet2012statistical}. If we use Eq.~(\ref{C-Eq}), the $z$-component of the velocity satisfy the Laplace equation 
\begin{equation}
\partial_\alpha \partial_\alpha v_{z} = 0.
\label{laplaceVzEq}
\end{equation}
On the other hand, the $z$-component of the vorticity takes the form
\[
\omega_z = \frac{1}{r} \left[\partial_r (r v_{\phi}) - \partial_\phi v_{r} \right] = \frac{1}{r} \left[ \partial_r (r v_{\phi}) - \frac{\nu}{2\Omega}\partial_r \partial_\phi \omega_z\right]. 
\] 
Since the vorticity is axially symmetric, then the previous relationship results in 
\begin{equation}
\omega_z = \frac{1}{r} \partial_r \left(r v_{\phi}\right). 
\label{D-Eq}
\end{equation}
Knowing that, the stationary vorticity equation from Eq.~(\ref{vorticityEq}) is
\begin{equation}
\partial_\beta\partial_\beta\omega_{\alpha} = -\frac{2\Omega}{\nu} \partial_z v_{\alpha}.
\end{equation}
In particular, the $z$-component is
\begin{equation}
\partial_\beta\partial_\beta \omega_{z} =  \frac{2\Omega}{\nu} \Pi.
\label{steadyzVorticityEq}
\end{equation}
As a result, the stationary velocity and pressure (in the framework of the highly condensed vorticity approximation) can be obtained with the following procedure: first, solve the Laplace equation given by Eq.~(\ref{laplaceVzEq}) for the $z$-component of velocity with appropriate boundary conditions. Second, compute $\Pi=-\partial_z v_z$ and solve the Poisson's equation Eq.~(\ref{steadyzVorticityEq}) to find the $z$-component of the vorticity. Third, use Eq.~(\ref{B-Eq}) to find the radial component of the velocity. Fourth, find the angular velocity from Eq.~(\ref{D-Eq}) by integrating of the $z$-component of $\vec{\omega}$. Finally, compute the pressure by solving Eq.~(\ref{PoissonForPWhenOmegaLargeEq}).

\section{The stationary z-velocity}
\label{The_stationary_z-velocitySection}
The $z$-component of the velocity is obtained by solving Eq.~(\ref{laplaceVzEq}). The Laplace equation in cylindrical coordinates is
\[
\vec{\nabla}^2 v_z = \frac{1}{r}\partial_r \left( r \partial_r v_z \right) +  \frac{1}{r^2}\partial_\phi \partial_\phi v_z + \partial_z\partial_z v_z = 0.
\]
This equation can be solved by using a standard method of separation of variables \cite{strauss2007partial,zill2012first}. The general solution can be written as
\begin{equation}
v_z(\vec{r}) = \sum_{m,n}\left[A_{mn} J_m(k_n r) + B_{mn} N_m(k_n r)\right]\exp(\pm \mathbf{i}m\phi)\exp(\pm k_n z) 
\label{generalSolutionEq}
\end{equation}
where $J_m(u)$ and  $N_m(u)$ are the Bessel functions of the first and second kind respectively. The series representation of these functions are
\[
J_m(u) = \sum_{n=0}^{\infty} \frac{(-1)^n (u/2)^{m+2n}}{n!\Gamma(m+n+1)}
\]
and
\[
N_m(u) = \frac{J_m(u)\cos(m\pi)-J_{-m}(u)}{\sin(m\pi)}.
\]

\subsection{Axially symmetric case $v_z=v_z(r,z)$} 
There are models for atmospheric flows where vorticity comes from the rotation of the earth where a vertical wind column of air generates the vortex \cite{leonov2014analytical}. Since wind has a small viscosity coefficient in comparison with incompressible fluids as water, then the diffusive term may vanish but the advective terms cannot. We study the opposite case where the fluid is highly viscous and motion sufficiently slow to drop the advective terms. However, it is possible to keep the idea where a vertical column of moving fluid can generate a vortex. For simplicity, the study will be focused on the axially symmetric case $v_z=v_z(r,z)$. For these type of flows one can consider without loss of generality that flow domain $\mathfrak{D}$ is a cylinder of radius $a$ and height $h$ centred at the origin $\mathfrak{D}:=[0,a] \times [0,\pi] \times [-h/2,h/2]$. We shall study the flow with the following Dirichlet boundary conditions: The $z$-velocity on the top and bottom bases located at $z=-h/2$ and $z=h/2$ are defined by the following functions    
\[
v_z(r,h/2) = f_{+}(r)\hspace{1.0cm}\mbox{and}\hspace{1.0cm}v_z(r,-h/2) = f_{-}(r).
\]
and $v_z(r,z)$ is set zero on the side of the cylinder
\[
v_z(a,z) = 0 \hspace{0.5cm}\mbox{for}\hspace{0.5cm}  |z| \leq h/2
\]
Since the flow can be considered axially symmetric, then $m=0$ because other modes do not contribute. The general solution given by Eq.~(\ref{generalSolutionEq}) takes the form
\[
v_z(\vec{r}) = \sum_{j=1}^{\infty} \left(v_z(r,z)\right)_{k_j} = \sum_{j=1}^{\infty} R(k_j r) Z(k_j z) 
\]
with
\[
 R(kr) = A J_o(k r)\hspace{1.0cm}\mbox{and}\hspace{1.0cm} Z(kz) = C_1 e^{k z} + C_2 e^{-k z} .
\]
Now, the boundary condition at $r=a$ implies that
\[
 \left(v_z(a,z)\right)_{k_j} = J_o(k_j a)Z(k_j z) = 0.
\]
Therefore $k_j=\alpha_j/a$ with $\alpha_j$ the jth-zero of $J_o(u)$. The solution may be written as follows 
\[
v_z(\vec{r}) = \sum_{j=1}^{\infty} \left(c_j^{(1)} e^{k_j z} + c_j^{(2)} e^{-k_j z}\right) J_o(\alpha_j r / a)
\]
with $c_j^{(1)}$ and $c_j^{(2)}$ constants. Those constants may be found by using the boundary condition on the top and bottom bases of the cylinder
\[
v_z(r, \pm h/2) = \sum_{j=1}^{\infty} \left(c_j^{(1)} e^{\pm k_j h/2} + c_j^{(2)} e^{\mp k_j h/2}\right) J_o(\alpha_j r / a) = f_{\pm}(r).
\]
Now, using the orthogonality condition of the Bessel functions
\[
 \int_{0}^a J_m(\alpha_{m,j}r/a) J_n(\alpha_{n,j}r/a) r dr = \frac{1}{2} a^2 [J_{m+1}(\alpha_{m,j})]^2 \delta_{mn}	
\]
the conditions at $z = \pm h/2$ may be written as 
\[
\left(
\begin{matrix}
  e^{+k_j h/2} & e^{-k_j h/2} \\
  e^{-k_j h/2} & e^{+k_j h/2}
 \end{matrix}
\right)\left(
\begin{matrix}
  c_j^{(1)} \\
  c_j^{(2)} 
 \end{matrix}
\right)
=
\left(
\begin{matrix}
  I_j\left\{f_{+}(r)\right\} \\
  I_j\left\{f_{-}(r)\right\} 
 \end{matrix}
\right)
\]
where we introduce the following integral function 
\[
I_j\left\{f(r)\right\} := \frac{2}{a^2 J_1^2(\alpha_j)} \int_{0}^\infty f(r) J_o(\alpha_j r/a) r dr
\]
which depends on the radial boundary functions. Solving the linear set of equations for the coefficients $c_j^{(1)}$ and $c_j^{(2)}$, we obtain
\begin{equation}
 c_j^{(1)} = \left[ e^{+k_j h/2} I_j\left\{f_{+}(r)\right\} - e^{-k_j h/2} I_j\left\{f_{-}(r)\right\} \right] / 2 \sinh(k_j h)
 \label{coefficient1Eq}
\end{equation}
and
\begin{equation}
c_j^{(2)} = \left[ e^{+k_j h/2} I_j\left\{f_{-}(r)\right\} - e^{-k_j h/2} I_j\left\{f_{+}(r)\right\} \right] / 2 \sinh(k_j h).  
\label{coefficient2Eq}
\end{equation}
\subsubsection{Gaussian profile} 
In the present, problem we suppose that $z$-component of velocity at the inlet and outlet ($z=\pm h/2$) behaves as  Gaussian functions 
\[
 v_z(r,\pm h/2) = f_{\pm}(r) = \frac{v_o a^2}{2\pi\sigma^2} \exp\left(-\frac{r^2}{2\sigma^2}\right) := f(r)  
\]
where $v_o$ has units of velocity and $\sigma$ is the standard deviation of the Gaussian distribution.
\begin{figure}[h]
  \centering   
  \includegraphics[width=0.5\textwidth]{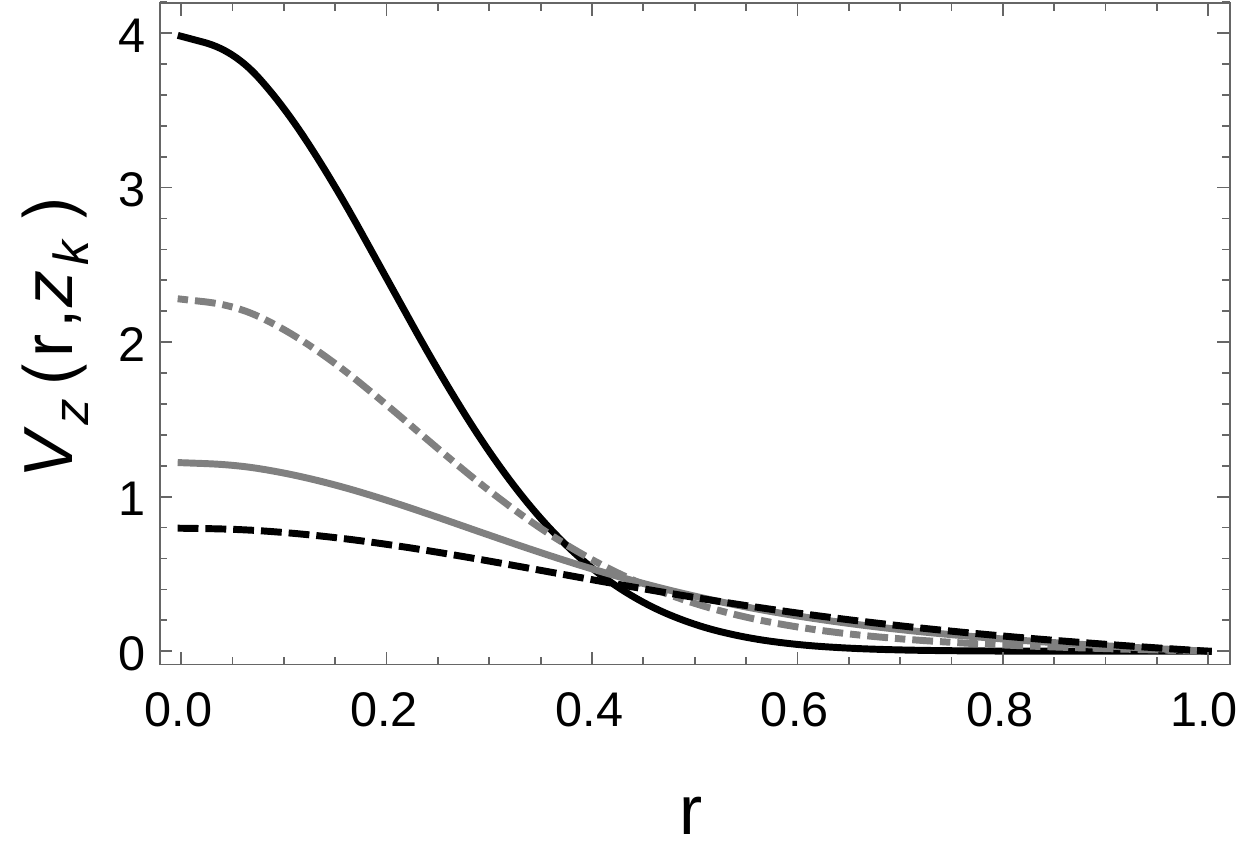}\hspace{0.2cm}
  \includegraphics[width=0.45\textwidth]{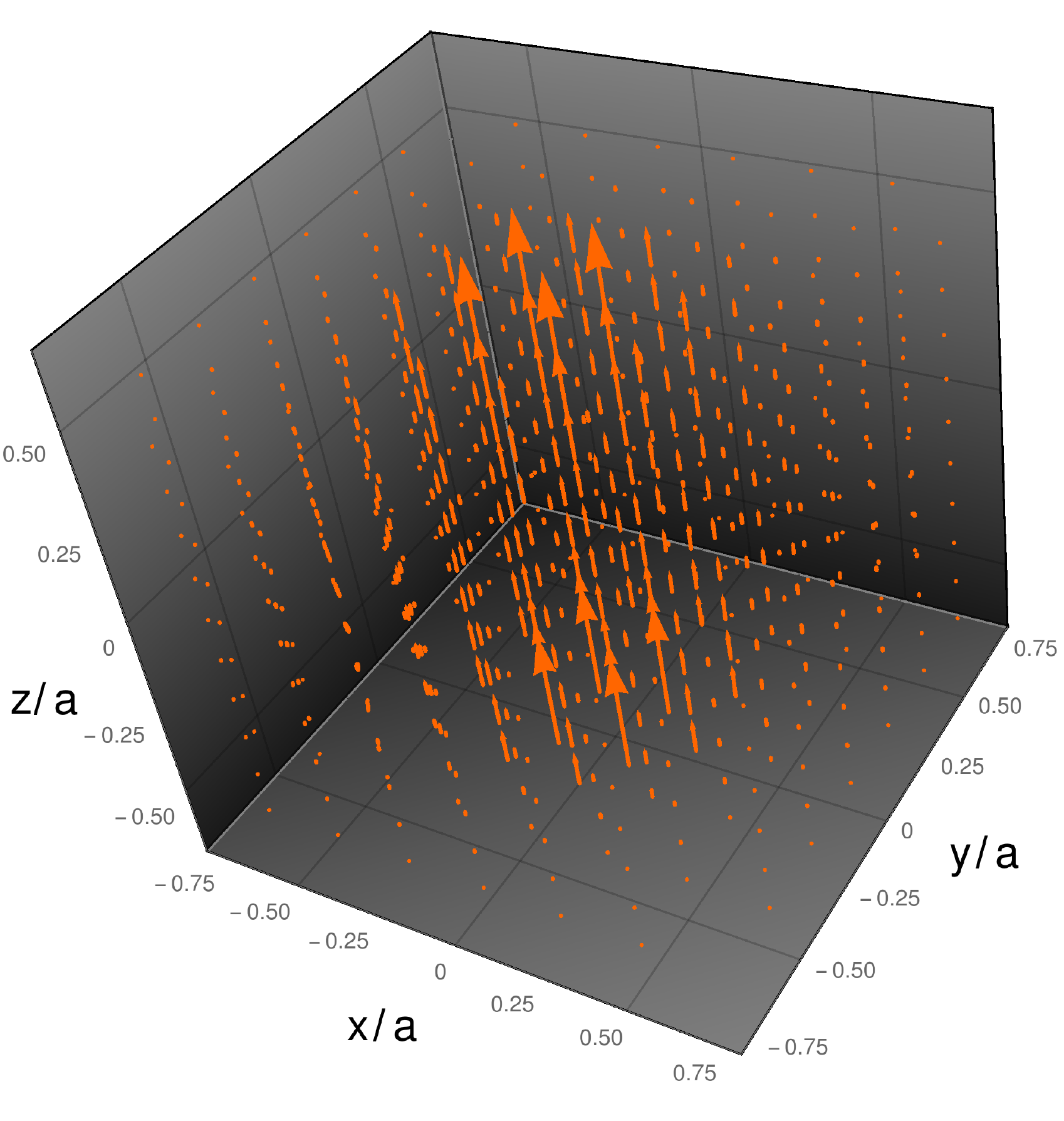}
  \caption[Vertical velocity.]%
  {$z$-th component of velocity. \textbf{(left)} Profile of $v_z(r,z)$ at different values of z. The black dashed line, gray solid line, gray dotdashed line and black solid line corresponds to z/a=0, 0.25, 0.4 and 0.5. \textbf{(right)} Vector plot of $v_z(r,z)$. This vertical column of moving fluid will generate the vortex.}
  \label{verticalVelocityFig}
\end{figure}\\
Since the inlet and outlet velocity profiles are the same, then the coefficients defined by Eqs.~(\ref{coefficient1Eq}) and (\ref{coefficient2Eq}) result in
\[
c_j^{(1,2)} = \frac{1}{2 \cosh(k_j h/2)} I_j\left\{f(r)\right\}.
\]
The integral $I_j\left\{f(r)\right\}$ can be found exactly for this type of Gaussian profile, giving
\[
 I_j\left\{f(r)\right\} = \frac{v_o a^2}{(\sigma a)^2 J_1^2(\alpha_j)} \int_{0}^\infty  \exp\left(-\frac{r^2}{2\sigma^2}\right) J_o(\alpha_j r/a) r dr = \frac{v_o}{\pi J_1^2(\alpha_j)} \exp\left[-\frac{1}{2}\left(\frac{\alpha_j\sigma}{a}\right)^2\right].
\]
Hence, the $z$-component of the velocity is found to be
\[
v_z(r,z) = \frac{v_o}{\pi}\sum_{j=1}^{\infty} \exp\left[-\frac{1}{2}\left(\frac{\alpha_j\sigma}{a}\right)^2\right] \frac{\cosh(\alpha_j z/a)}{\cosh(\alpha_j h/(2a))}  \frac{J_o(\alpha_j r/a)}{J_1^2(\alpha_j)}
\]
and
\[
\Pi(r,z) = -\frac{v_o}{\pi a}\sum_{j=1}^{\infty} \exp\left[-\frac{1}{2}\left(\frac{\alpha_j\sigma}{a}\right)^2\right] \frac{\sin(\alpha_j z/a)}{\cosh(\alpha_j h/(2a))} \alpha_j \frac{J_o(\alpha_j r/a)}{J_1^2(\alpha_j)}.
\]
A plot of the $z$-velocity is shown in Fig.~\ref{verticalVelocityFig} where the parameters $\sigma=0.2$, $v_o$, $h$ and $a$ have been set as one. In principle, the vertical motion of the fluid (see Fig.~\ref{verticalVelocityFig}-(right)) will generate the vortex via the function $\Pi(r,z)$ since the Poisson equation for the vorticity depends on it. Additionally, $\Pi(r,z)$ is also an odd function
\begin{equation}
\Pi(r,-z) = -\Pi(r,z)     
\label{parityConditiionOnPiEq}
\end{equation}
because $\sinh(-z) = -\sinh(z)$. This parity property of $\Pi(r,z)$ will be inherited by the vorticity affecting the dynamics of the flow as it will be described in the following sections.

\section{Integral form of the vorticity}
\label{Integral_form_of_the vorticitySection}
Let now consider the solution of Eq.~(\ref{steadyzVorticityEq}) in a cylindrical region $\mathfrak{D}\in \mathcal{R}^3$ specifying Dirichlet boundary conditions on $S$. This problem is analogous to find the electric potential $V(\vec{r})$ of an stationary charge distribution $\rho(\vec{r})$ where the $z$-vorticity and $\Pi(\vec{r})$ play the role of electric potential and density charge respectively. We may use this analogy between steady fluid flow and electrostatics to write the general integral solution of the problem as follows 
\begin{equation}
\omega_z(\vec{r})=-\frac{\Omega}{2\pi\nu}\int_{\mathfrak{D}} G(\vec{r},\pvec{r}') \Pi(\pvec{r}') d^3\pvec{r}' - \frac{1}{4\pi}\oint_S \omega_z(\pvec{r}') n'_{\alpha} \partial'_{\alpha} G(\vec{r},\pvec{r}') dS'
\label{omegazIntegralSolEq}
\end{equation}
where $G(\vec{r},\pvec{r}')$ is the green function \cite{eisler1969introduction,stakgold2011green} of the Laplacian in the three-dimensional space
\[
G(\vec{r},\pvec{r}') = \frac{1}{|\vec{r}-\pvec{r}'|}
\]  
which implies 
\[
\partial_\alpha \partial_\alpha G(\vec{r},\pvec{r}') = -4\pi\delta(\vec{r}-\pvec{r}').
\]
In general the surface integral of Eq.~(\ref{omegazIntegralSolEq}) vanishes if bounding surface goes out to infinity. If the function $\Pi(\pvec{r}')$ is concentrated near the $z$-axis, then the vorticity will decrease as $r$ increases and the surface integral term on the walls of $S$ will vanish. However, the contribution of the surface integral at the top and bottom bases of the cylindrical domain is not necessarily zero because the parity of $\Pi(\pvec{r}')$ introduces a vorticity inversion along the $z$-axis. It implies that inlet and outlet vorticities would have opposite sign $\omega_z(r',h/2) = -\omega_z(r',-h/2)$ if $f_{-}(r)=f_{+}(r)$. We shall solve the problem assuming that bounding surface goes out to infinity, then   
\begin{equation}
\omega_z(\vec{r})=-\frac{\Omega}{2\pi\nu}\int_{\mathbb{R}^3} G(\vec{r},\pvec{r}') \Pi(\pvec{r}') d^3\pvec{r}' 
\label{omegazIntegralSolTwoEq}
\end{equation}
where $\Pi(\pvec{r}')$ is thought as a source or sink of vorticity whose value is zero if $\pvec{r}'\notin\mathfrak{D}$. In cylindrical coordinates, it is convenient to expand the Green's function as follows \cite{cohl1999compact}
\[
\frac{1}{|\vec{r}-\pvec{r}'|} = \frac{1}{\pi \sqrt{r r'}} \sum_{m=-\infty}^\infty \exp\left[\mathbf{i}m(\phi-\phi')\right]Q_{m-1/2}\left[\chi(\vec{r},\pvec{r}')\right]
\]
where
\[
\chi(\vec{r},\pvec{r}') := \frac{1}{2 r r'} \left[r^2 - r'^2 + (z-z')^2\right]
\]
and $Q_{m-1/2}$ is the half-integer degree Legendre function of the second kind. The compact expansion of the Green's function may be written as follows
\begin{equation}
\frac{1}{|\vec{r}-\pvec{r}'|} = \frac{1}{\pi \sqrt{r r'}} \sum_{m=0}^\infty 2\cos(m(\phi-\phi'))Q_{m-1/2}\left[\chi(\vec{r},\pvec{r}')\right] + \frac{1}{\pi \sqrt{r r'}} Q_{-1/2}\left[\chi(\vec{r},\pvec{r}')\right]
\label{GreensCompactExpansionIIEq}
\end{equation}
due to the even parity of the $Q_{m-1/2}$. Since $\Pi(\pvec{r}')$ does not depends on $\phi$, then only the $m=0$ term of Eq.~(\ref{GreensCompactExpansionIIEq}) will contribute to the vorticity integral Eq.~(\ref{omegazIntegralSolTwoEq})
\[
\omega_z(\vec{r})=-\frac{\Omega}{\pi\nu \sqrt{r}}\int_{\Xi} dr'dz' \sqrt{r'}\Pi(r',z') Q_{-1/2} \left[\chi(\vec{r},\pvec{r}')\right].
\]
where the integral is over all the $rz$-plane. A useful expression to evaluate $Q_{m-1/2}$ is \cite{abramowitz1965handbook}
\[
Q_{-1/2} \left[\chi(\vec{r},\pvec{r}')\right] = \mu(\vec{r},\pvec{r}') K[\mu(\vec{r},\pvec{r}')]
\]
where $K(\mu)$ is the \textit{elliptic integral of the first kind} defined as follows
\begin{equation}
    K(\chi) = \int_{0}^{\pi/2} \frac{d\phi}{1-\chi^2\sin^2\phi}
\label{KIntegralDefinitionEq}
\end{equation}
and 
\[
\mu(\vec{r},\pvec{r}') := \sqrt{\frac{2}{1+\chi}} = \sqrt{\frac{4 r r'}{(r+r')^2 + (z-z')^2}}.
\]
Therefore the $z$-vorticity takes the form
\begin{equation}
\omega_z(\vec{r}) = \frac{\Omega}{\pi\nu}\int_{\Xi} dr'dz' \sqrt{r'} \left.\left( \partial_z v_z \right)\right|_{(r',\phi')} \frac{\mu(\vec{r},\pvec{r}') }{\sqrt{r}}K[\mu(\vec{r},\pvec{r}')] .
\label{vorticityFirstIntegralSoluitonEq}
\end{equation}

\section{Point-wise distribution approximation for the $\Pi$ function}
\label{Point-wise_distribution_approximation_for_Pi_functionSection}
The main difficulty on the computation of the integral solution given by Eq.~(\ref{vorticityFirstIntegralSoluitonEq}) comes from the complexity of $\Pi(\vec{r})$ as occurs with the charge density in the electrostatic counterpart. Then, a strategy to evaluate Eq.~(\ref{vorticityFirstIntegralSoluitonEq}) is to approximate the function $\Pi(\vec{r})=-\partial_z v_z$ as it would be a \textit{distribution of point-like sinks/sources of vorticity} in $\mathfrak{D}$. 
\begin{figure}[h]
  \centering   
  \includegraphics[width=0.7\textwidth]{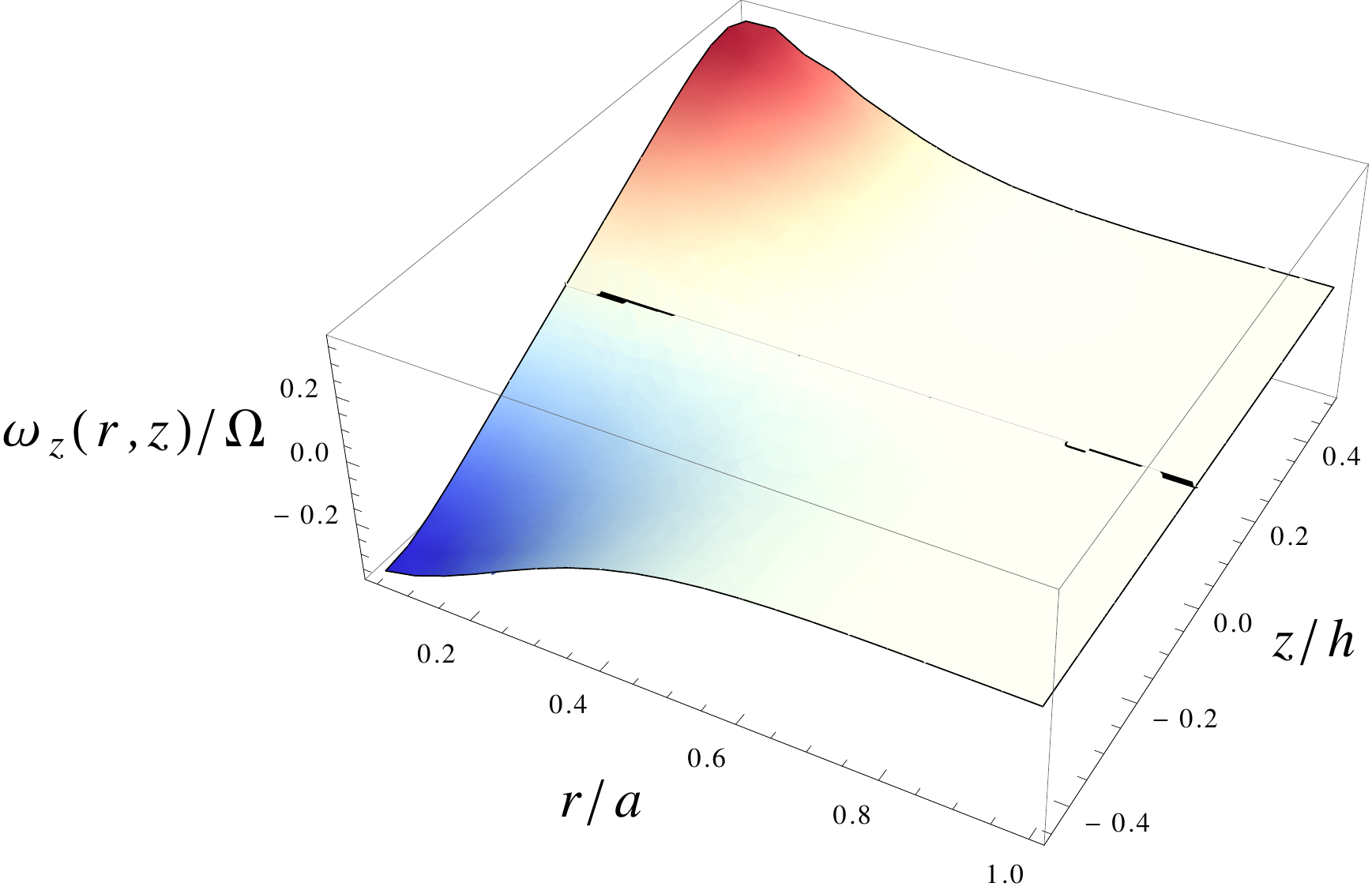}
  \caption[Vorticity.]%
  {\textbf{Vorticity.} Approximated profile of $\omega_z$ according to Eq.~(\ref{vorticityPunctualChargeApproximationEq}) with $N=400$ point-like sinks/sources of vorticity distributed uniformly on the plane $\Upsilon$.}
  \label{voriticityFig}
\end{figure}\\
Since $\Pi(\vec{r})$ can be written as
\[
\Pi(\vec{r}) = \int_{\mathbb{R}^3} d^3\pvec{r}' \delta(\vec{r}-\pvec{r}')\left.\left(-\partial_z v_z\right)\right|_{\pvec{r}'} = \int_{0}^{2\pi} d\phi' \int_{0}^{\infty} r' dr' \int_{-\infty}^{\infty} dz' \frac{\delta(r-r')}{2\pi r'} \delta(z-z')\delta(\phi-\phi')\left.\left(-\partial_z v_z\right)\right|_{\pvec{r}'},
\]
then
\[
\Pi(r,z) = \int_{\Xi} dr'dz' \delta(r-r')\delta(z-z')\left.\left(-\partial_z v_z\right)\right|_{(r',\phi')},
\label{piFunctionAgainEq}
\]   
by virtue of the azimuthal symmetry of $\Pi$. We approximate Eq.~(\ref{piFunctionAgainEq}) as follows\footnote{This formula is analogous to \[\rho_q(\vec{r}) = \sum_{i=1}^N q_i\delta(\vec{r}_i-\vec{r})\] where $\rho_q(\vec{r})$ is the density distribution  of a set of discrete charges $q_1,\ldots,q_N$ located at $\vec{r}_1,\ldots,\vec{r}_N$.  }  
\[
\Pi(r,z) = \lim_{N\to\infty} \sum_{\vec{r}_{\alpha} \in \mathcal{M}_N} \tilde{q}_\alpha \delta(r-r_{\alpha})\delta(z-z_{\alpha})
\]
where $\Upsilon=\left\{(r,z) : 0 \leq r \leq a \wedge |z| \leq h/2 \right\}\in\mathbb{R}^2$,
\[
\tilde{q}_{\alpha} := \frac{ah}{N} \left.\left(-\partial_z v_z\right)\right|_{\vec{r}=\vec{r}_{\alpha}}  
\]
is playing the role of the $\alpha$-th punctual sink/source of vorticity, $\alpha = 1,\ldots,N$ and $N$ is the total number of point-like vorticity carriers with positions in $\mathcal{M}_N = \left\{\vec{r}_1,\ldots,\vec{r}_N\right\} \in \Upsilon$. Therefore the vorticity is written as follows
\begin{equation}
\omega_z(\vec{r}) = -\frac{\Omega}{\pi\nu}\lim_{N\to\infty} \sum_{\vec{r}_{\alpha} \in \mathcal{M}_N} \sqrt{r_{\alpha}} \tilde{q}_\alpha \frac{\mu(\vec{r},\vec{r}_{\alpha}) }{\sqrt{r}}K[\mu(\vec{r},\vec{r}_{\alpha})].
\label{vorticityPunctualChargeApproximationEq}
\end{equation}
We use the previous formula in order to compute the $z$-component of the vorticity taking finite, but large, values of $N$. The vorticity distribution obtained with this approximation is shown Fig.~\ref{voriticityFig}. Note that $\tilde{q}_{\alpha}=\tilde{q}(r_{\alpha}, z_{\alpha})$ does not depend on $\phi$ because it belongs to $\Upsilon$. Moreover, these punctual carries of vorticity inherit the parity property of Eq.~(\ref{parityConditiionOnPiEq}), this is $\tilde{q}(r_{\alpha}, -z_{\alpha}) = - \tilde{q}(r_{\alpha}, z_{\alpha})$. Since the terms $\mu(\vec{r},\vec{r}_{\alpha})K[\mu(\vec{r},\vec{r}_{\alpha})]$ are even functions with respect the $z$-coordinate, then the change of $\textbf{sgn}(\omega_z)$ along the $z$-axis depends only on the distribution of the punctual sources $\{\tilde{q}_{\alpha}\}_{\alpha=1,\ldots,N}$ on $\Upsilon$. If the $z$-velocity inlet/outlet is a Gaussian profile, then the majority of vorticity carries in the region $z<0$ are sources of vorticity, and the $z$-vorticity in the bottom region is positive. The opposite situation occurs in the region $z>0$ where vorticity is negative. In general, the evaluation of Eq.~(\ref{vorticityPunctualChargeApproximationEq}) requires a uniform distribution of the punctual sinks/sources positions on $\Upsilon$, then $\sum_{\vec{r}_{\alpha} \in \mathcal{M}_N} \tilde{q}_\alpha = 0$ due to the inlet/outlet definition of the boundary conditions for the $z$-velocity. Then, the main consequence of this \textit{global neutrality} condition for symmetric distributions of $v_z$ along the $z$-axis (as the one shown in Fig.~\ref{verticalVelocityFig}-(right)) is the cancellation of the $z$-vorticity at $z=0$. 

\subsection{Radial velocity}
\label{Radial_velocitySubsection}
The next step of the procedure is to find the radial component of velocity. This may be found by replacing Eq.~(\ref{vorticityPunctualChargeApproximationEq}) in Eq.~(\ref{B-Eq}) obtaining 
\[
v_r(r,z) = \frac{1}{2\pi}\lim_{N\to\infty} \sum_{\vec{r}_{\alpha} \in \mathcal{M}_N} \sqrt{\vec{r}_{\alpha}} \tilde{q}_\alpha \partial_r\left\{\frac{\mu(\vec{r},\vec{r}_{\alpha}) }{\sqrt{r}}K[\mu(\vec{r},\vec{r}_{\alpha})]\right\}.
\]
We use the following derivative formula of the elliptic integral of the first kind
\[
\partial_\chi K(\chi) = \frac{E(\chi)-(1-\chi)K(\chi)}{2(1-\chi)\chi}
\]
to write 
\[
\partial_r\left\{\frac{\mu(\vec{r},\vec{r}_{\alpha}) }{\sqrt{r}}K[\mu(\vec{r},\vec{r}_{\alpha})]\right\} = \frac{\mu}{\sqrt{r}}\frac{1}{r}\left[\frac{r_{\alpha}^2-r^2+(z-z_\alpha)^2}{(r-r_{\alpha})^2+(z-z_\alpha)^2}E(\mu)-K(\mu)\right]
\]
where $E(\chi)$ is \textit{the complete elliptic integral of the second kind}, defined as
\begin{equation}
E(\chi) = \int_0^{\pi/2} \sqrt{1-\chi^2\sin^2\phi}d\phi.
\label{EIntegralDefinitionEq}    
\end{equation}
Therefore, the radial velocity takes the form
\begin{equation}
v_r(r,z) = \frac{1}{2\pi}\lim_{N\to\infty} \sum_{\vec{r}_{\alpha} \in \mathcal{M}_N} \sqrt{\vec{r}_{\alpha}} \tilde{q}_\alpha \frac{\mu(\vec{r},\vec{r}_{\alpha})}{r^{3/2}}\left\{\frac{r_{\alpha}^2-r^2+(z-z_\alpha)^2}{(r-r_{\alpha})^2+(z-z_\alpha)^2}E[\mu(\vec{r},\vec{r}_{\alpha})]-K[\mu(\vec{r},\vec{r}_{\alpha})]\right\}.
\label{radialVelocitySolutionWithDivergenceAtTheOriginEq}
\end{equation}
Note that Eq.~(\ref{radialVelocitySolutionWithDivergenceAtTheOriginEq}) has an inverse power dependence $1/(r^{3/2})$ term which diverges as $r \rightarrow 0$. On the other hand, the function $\mu(\vec{r},\vec{r}_\alpha)$ tends zero if $\vec{r}\neq\vec{r}_\alpha$ therefore     
\begin{equation}
\lim_{r \to 0} \left\{\frac{r_{\alpha}^2-r^2+(z-z_\alpha)^2}{(r-r_{\alpha})^2+(z-z_\alpha)^2}E[\mu(\vec{r},\vec{r}_{\alpha})]-K[\mu(\vec{r},\vec{r}_{\alpha})]\right\} = 0 
\label{limitAuxiliaryTermEq}    
\end{equation}
since $K(0)=E(0)=\pi/2$ (see Eqs.~(\ref{KIntegralDefinitionEq}) and (\ref{EIntegralDefinitionEq})).
\begin{figure}[h]
  \centering   
  \includegraphics[width=0.49\textwidth]{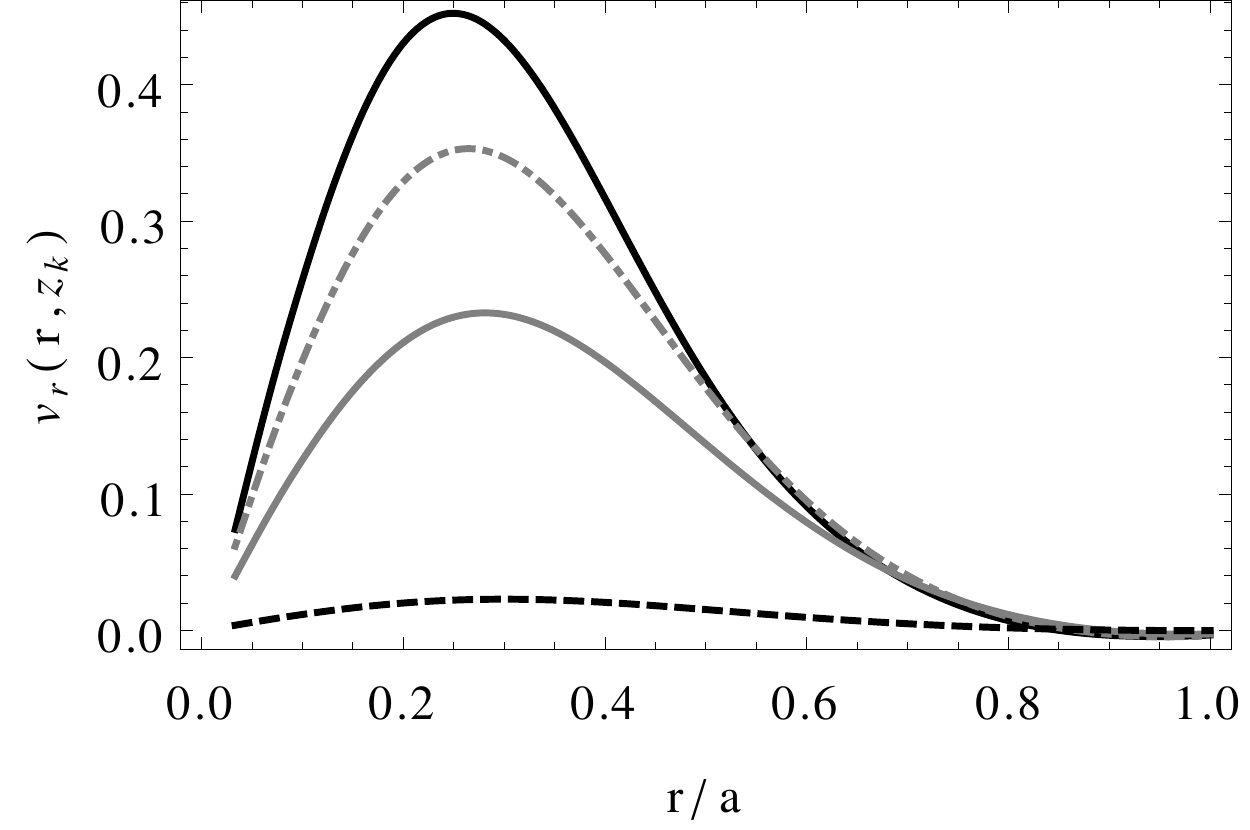}\hspace{0.1cm}
  \includegraphics[width=0.49\textwidth]{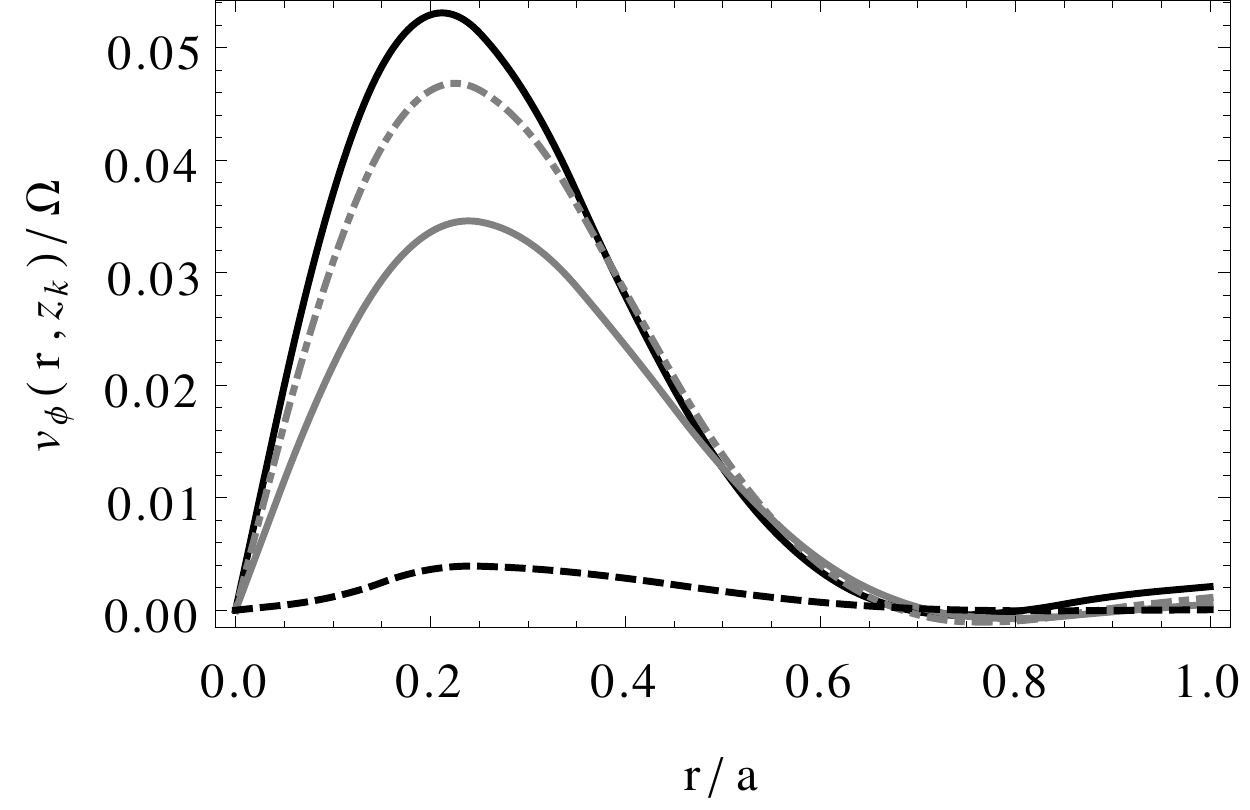}
  \caption[Vertical velocity.]%
  {Radial and angular velocities. \textbf{(left)} radial velocity and \textbf{(right)} tangent velocity at different values of $z$. The black dashed line, gray solid line, gray dot-dashed line and black solid line corresponds to $z/a$=0.023, 0.2234, 0.3234 and 0.4234.}
  \label{rAndPhiVelocitiesFig}
\end{figure}\\
In principle, there is no reason to find a divergent radial velocity at the origin since the $z$-velocity is not divergent and the continuity equation should avoid this situation. Then, we should expect that the inverse power divergence in Eq.~(\ref{radialVelocitySolutionWithDivergenceAtTheOriginEq}) must be cancelled by the term in Eq.~(\ref{limitAuxiliaryTermEq}). It is possible to handle this problem eliminating the $1/(r^{3/2})$ dependence by using the series representation of the elliptic integrals \cite{radon1950sviluppi}
\[
E(\chi) = \frac{\pi}{2} + \frac{\pi}{2}\sum_{m=1}^\infty e_m \chi^{2m} \hspace{0.5cm}\mbox{and}\hspace{0.5cm} K(\chi) = \frac{\pi}{2} + \frac{\pi}{2}\sum_{m=1}^\infty \kappa_m \chi^{2m} 
\]
where the coefficients $e_m$ and $\kappa_m$ are given, respectively, by
\[
e_m = \left[\frac{(2m-1)!!}{(2m)!!}\right]^2\frac{1}{1-2m} \hspace{0.5cm}\mbox{and}\hspace{0.5cm} \kappa_m = \frac{e_m}{1-2m}
\]
with $n!!$ the double factorial
\[ 
n!! = \begin{cases} 
      \prod_{k=1}^{n/2} (2k) & n>0 \mbox{ even}\\
      \prod_{k=1}^{(n+1)/2} (2k-1) & n>0 \mbox{ odd}\\
      1 & n=0
   \end{cases} .
\]
Finally, the radial velocity may be written as follows
\begin{eqnarray}
v_r(r,z) &=& \frac{1}{2\pi}\lim_{N\to\infty} \sum_{\vec{r}_{\alpha} \in \mathcal{M}_N} \sqrt{\vec{r}_{\alpha}} \tilde{q}_\alpha f(\vec{r}, \vec{r}_{\alpha})/4 \cdot \nonumber \\
         &\cdot& \left\{  \frac{r_\alpha^2-r^2+(z+z_\alpha)^2}{\sqrt{(r+r_\alpha)^2 + (z+z_\alpha)^2}}\sum_{m=1}^{\infty} e_m f(\vec{r}, \vec{r}_{\alpha})^m r^{m-1} \right. \nonumber \\ 
          &-& \left.\frac{2(r+r_\alpha)^2}{\sqrt{(r+r_\alpha)^2 + (z+z_\alpha)^2}}\sum_{m=1}^{\infty} \kappa_m f(\vec{r}, \vec{r}_{\alpha})^m r^{m-1} \right\}\frac{\pi}{2}
\end{eqnarray}
with
\[
f(\vec{r}, \vec{r}_{\alpha}) = \frac{r_\alpha}{\sqrt{(r+r_\alpha)^2 + (z+z_\alpha)^2}}.
\]
\begin{figure}[h]
  \centering   
  \includegraphics[width=0.25\textwidth]{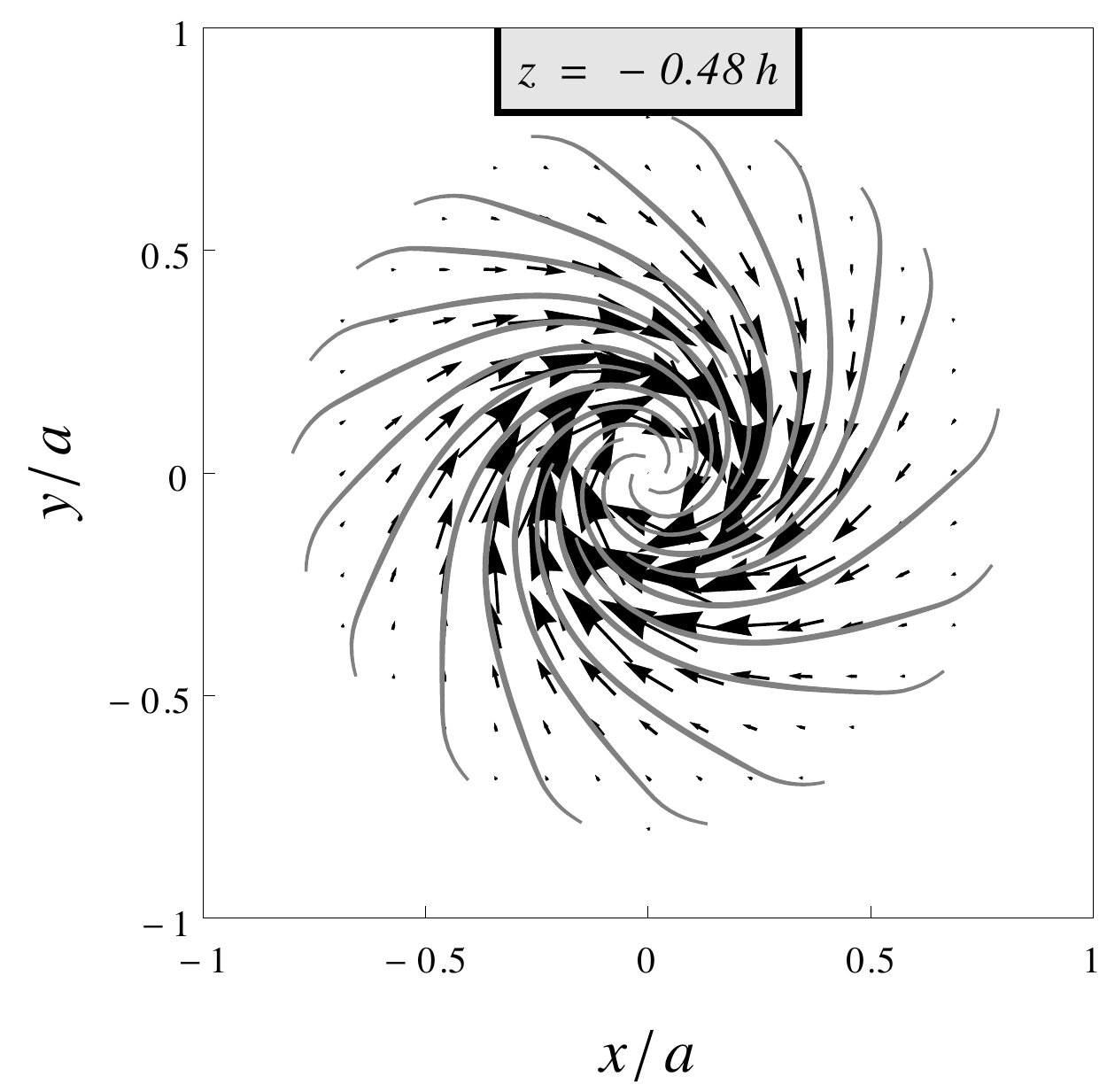}
  \includegraphics[width=0.25\textwidth]{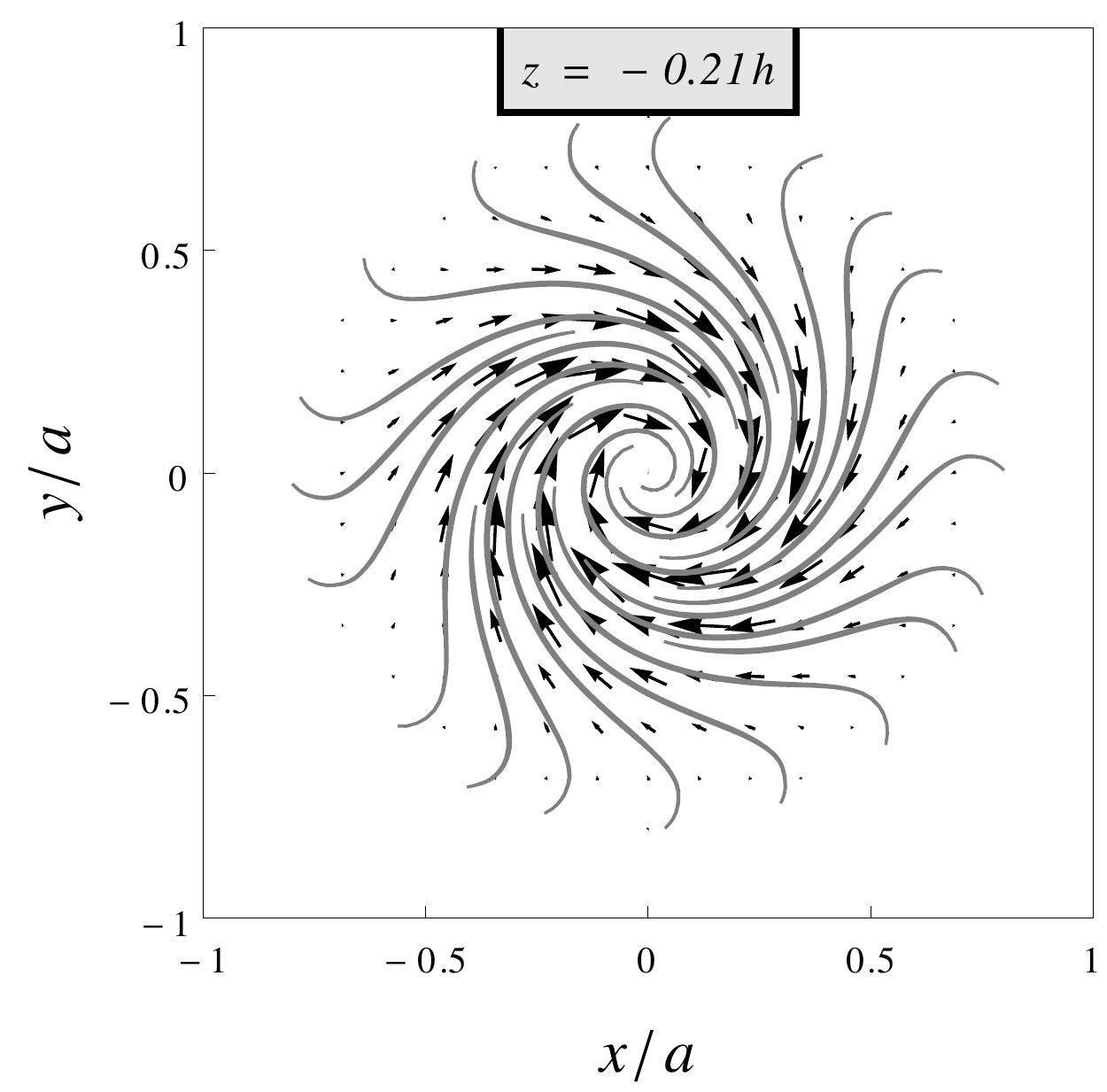}
  \includegraphics[width=0.25\textwidth]{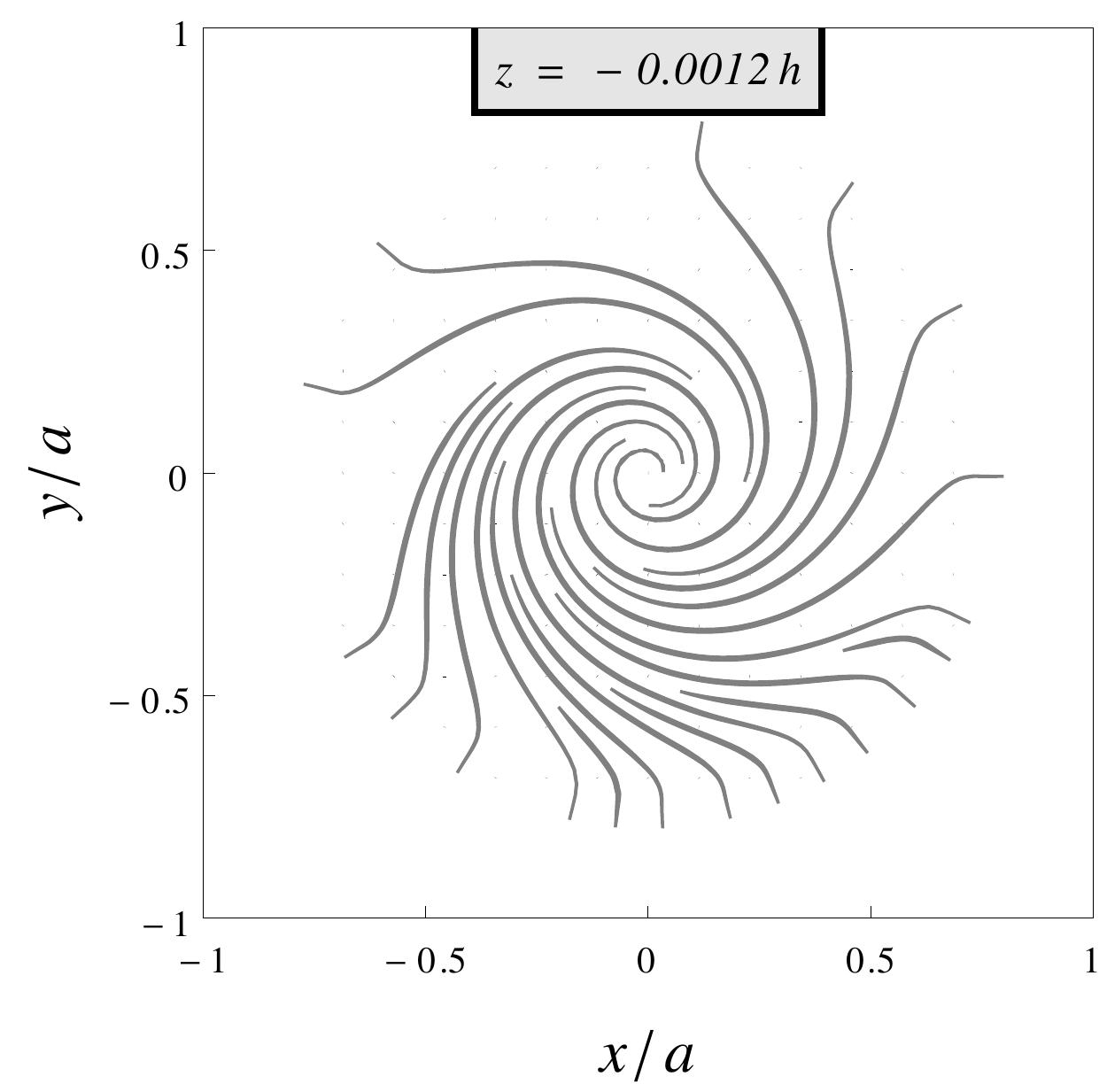}\\
  \includegraphics[width=0.25\textwidth]{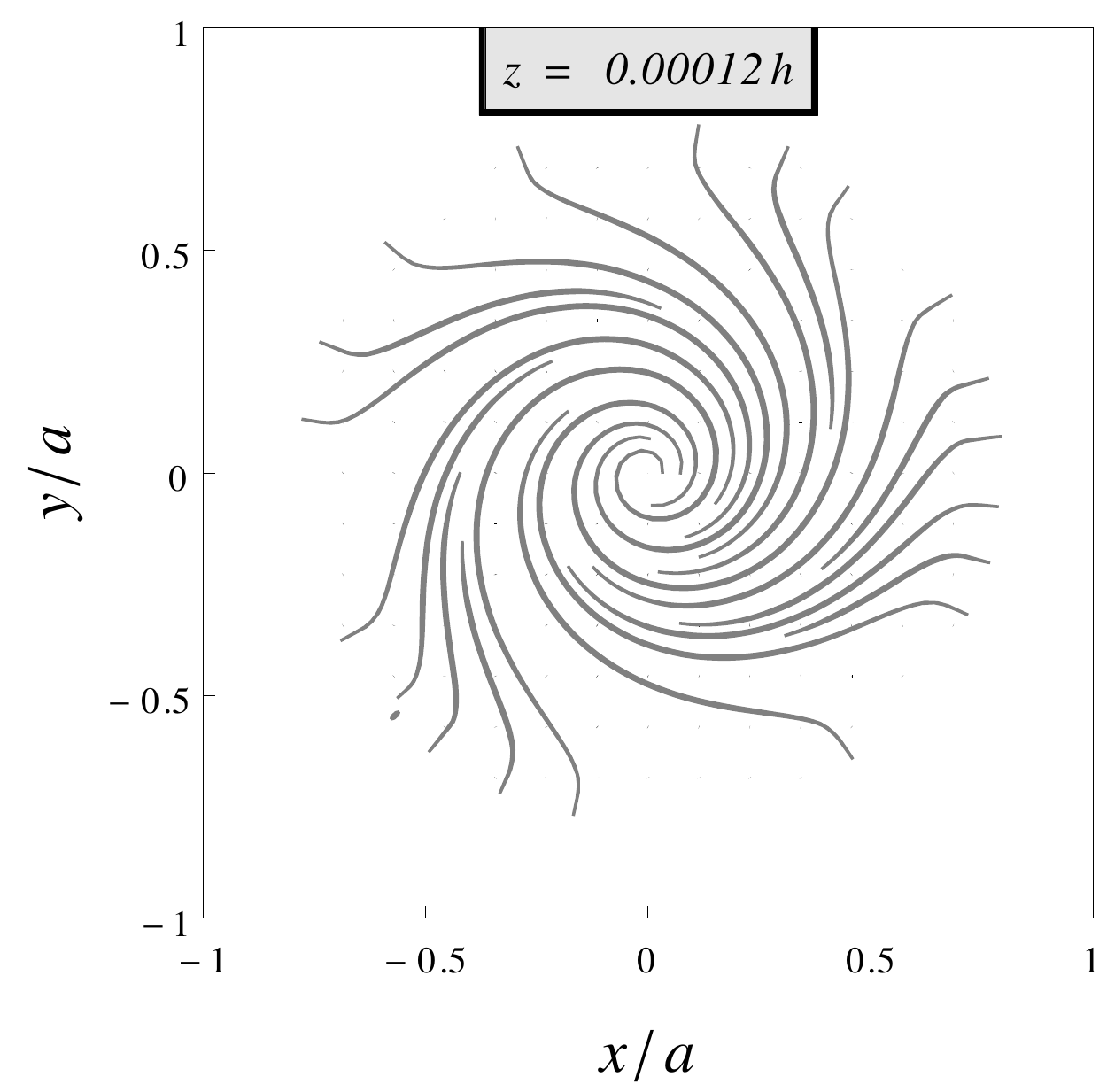}
  \includegraphics[width=0.25\textwidth]{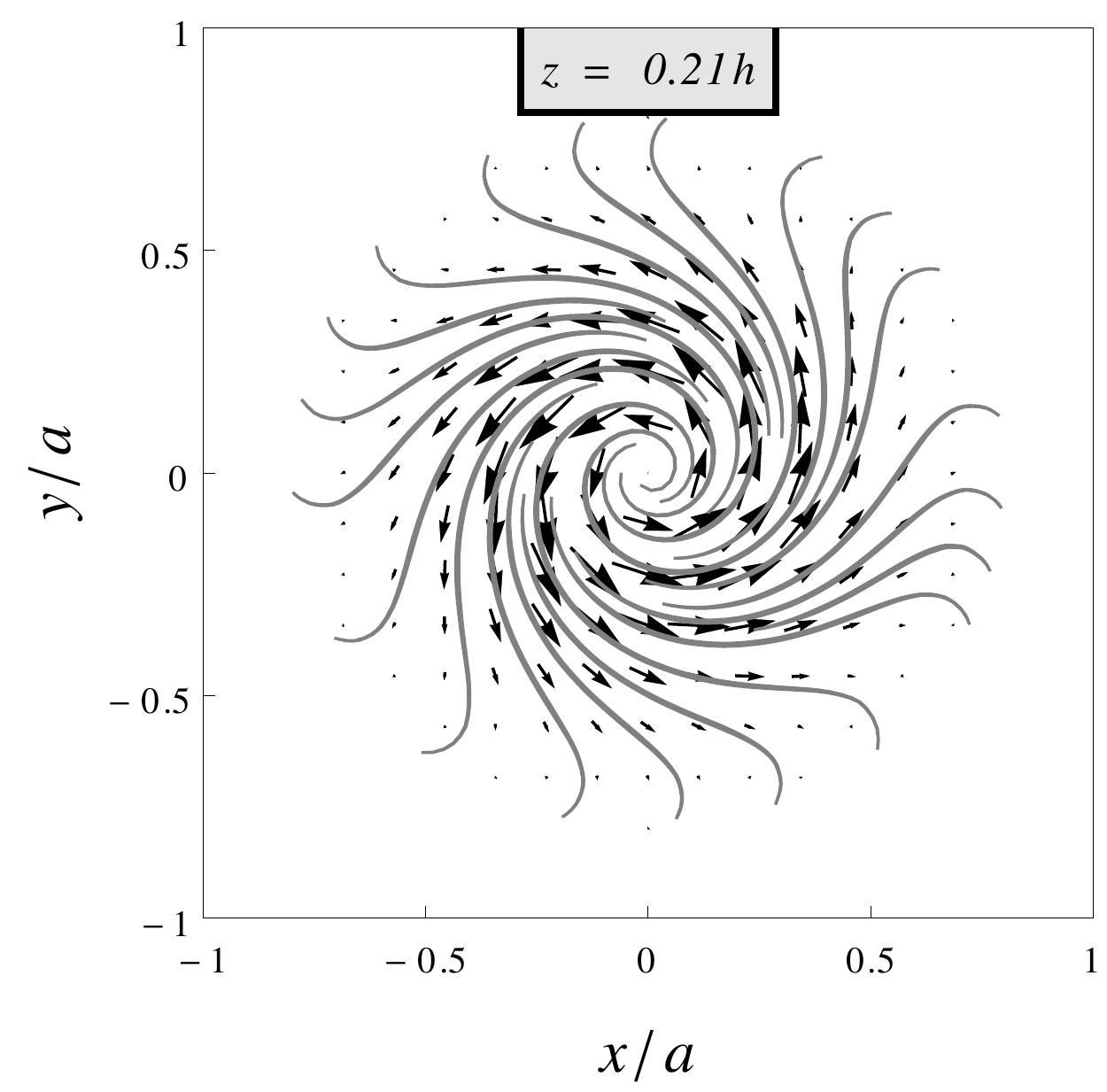}
  \includegraphics[width=0.25\textwidth]{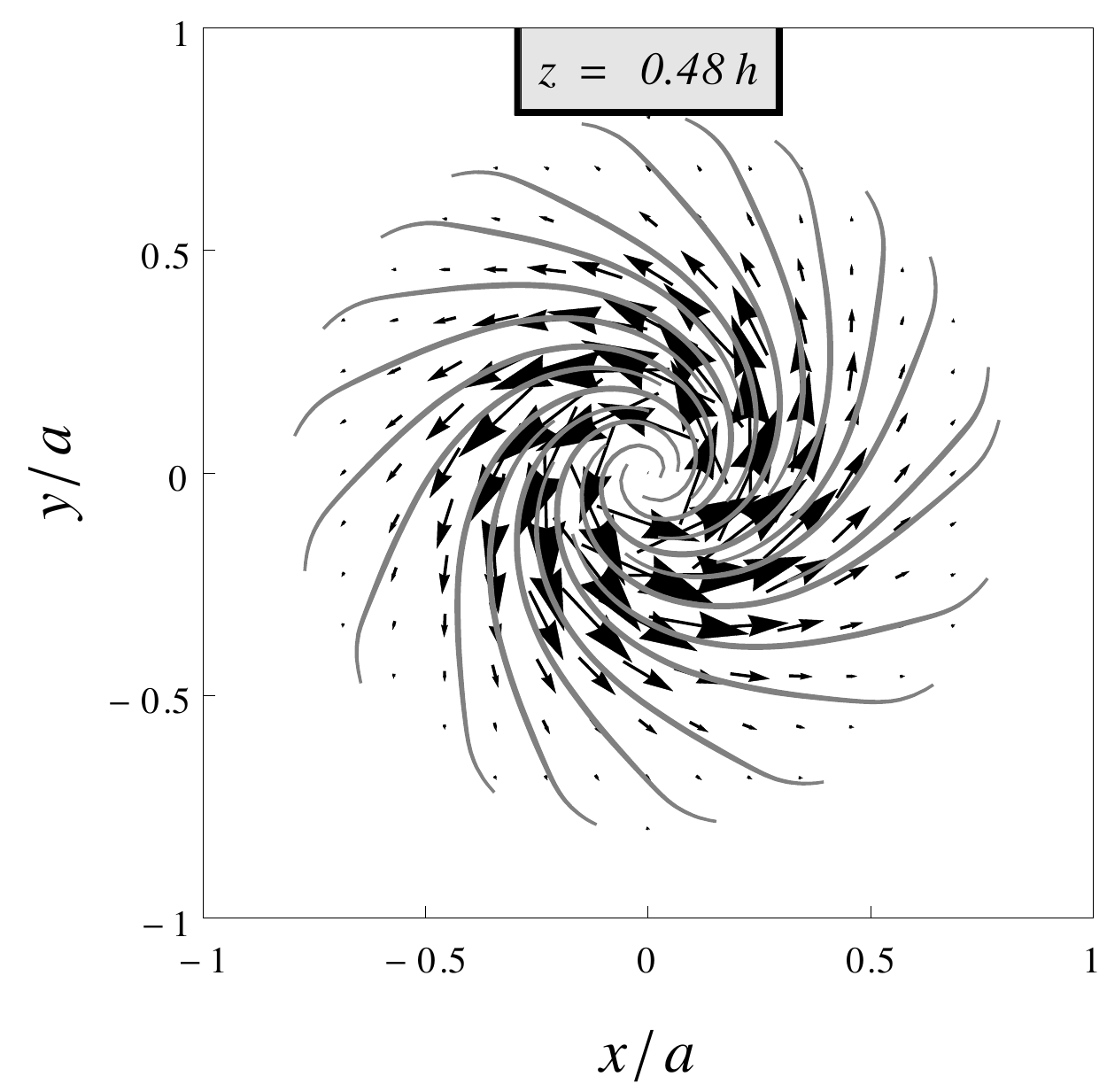}
  \caption[Velocity.]%
  {\textbf{Velocity.} Projection of the velocity field on planes located at $z/h=$ constant. }
  \label{velocitySectionsFig}
\end{figure}\\
A plot of the radial velocity profile at different values of $z/a > 0$ is shown in Fig.~\ref{rAndPhiVelocitiesFig}. The radial velocity and the $z$-vorticity share the same parity along the $z$-axis, this is $v_r(r, - z) = - v_r(r, - z)$. Then the fluid enters to the cylindrical region when $z<0$ since $v_r(r,z)<0$, and it exits $v_r(r,z)>0$ when $z>0$. The radial velocity is zero at $z=0$ as occurs with the $z$-vorticity. The extreme values of the radial velocity for $z/a = 0.4234, 0.3234 \mbox{ and } 0.2234$ are located at $z^{*}/a = 0.024 \pm 0.01, 0.026 \pm 0.01 \mbox{ and } 0.027 \pm 0.01$. Then there is a shifting to the right of the extreme value of $v_r$ as $|z|$ decreases.          

\subsection{Angular velocity}
\label{Angular_velocitySubSection}
It is possible to find the $\phi$-velocity by direct integration of Eq.~(\ref{D-Eq}) 
\[
r v_\phi(r,\phi,z) = \int_{0}^r r'\omega_z(r',z)dr + c(z,\phi)
\]
where $c(z,\phi)$ a function independent of the radial coordinate which may be found from the condition
\[
c(z,\phi)=\lim_{r \to 0} r v_\phi(r,\phi,z)=0.
\]
\begin{figure}[h]
\begin{center}
\begin{minipage}{0.95\linewidth}
\begin{center}
    \begin{minipage}{.5\linewidth}
        \centering
        \includegraphics[width=1.0\textwidth]{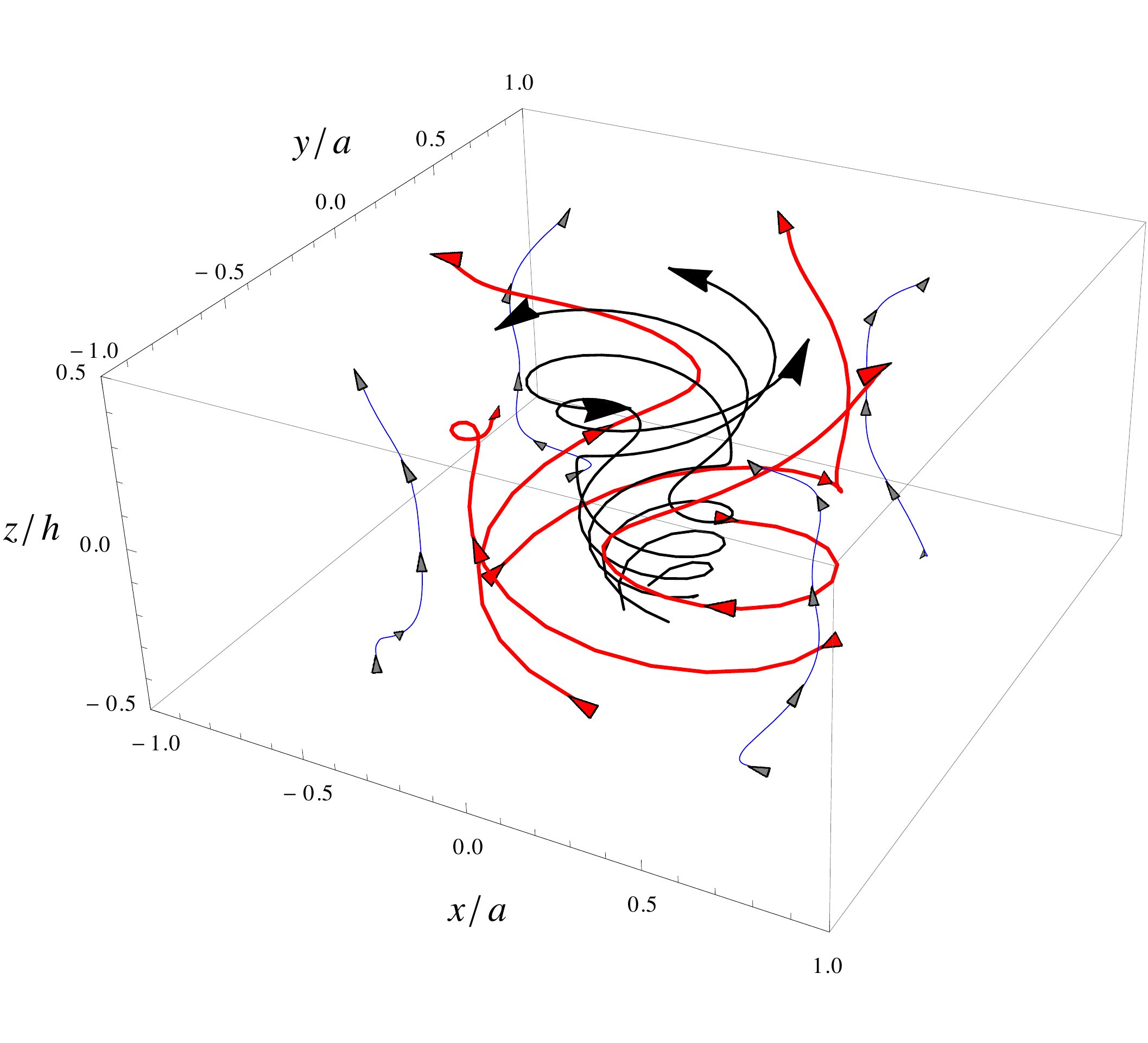}
        (a) stream lines.
    \end{minipage}\\
    \begin{minipage}{.3\linewidth}
        \centering
        \includegraphics[width=\textwidth]{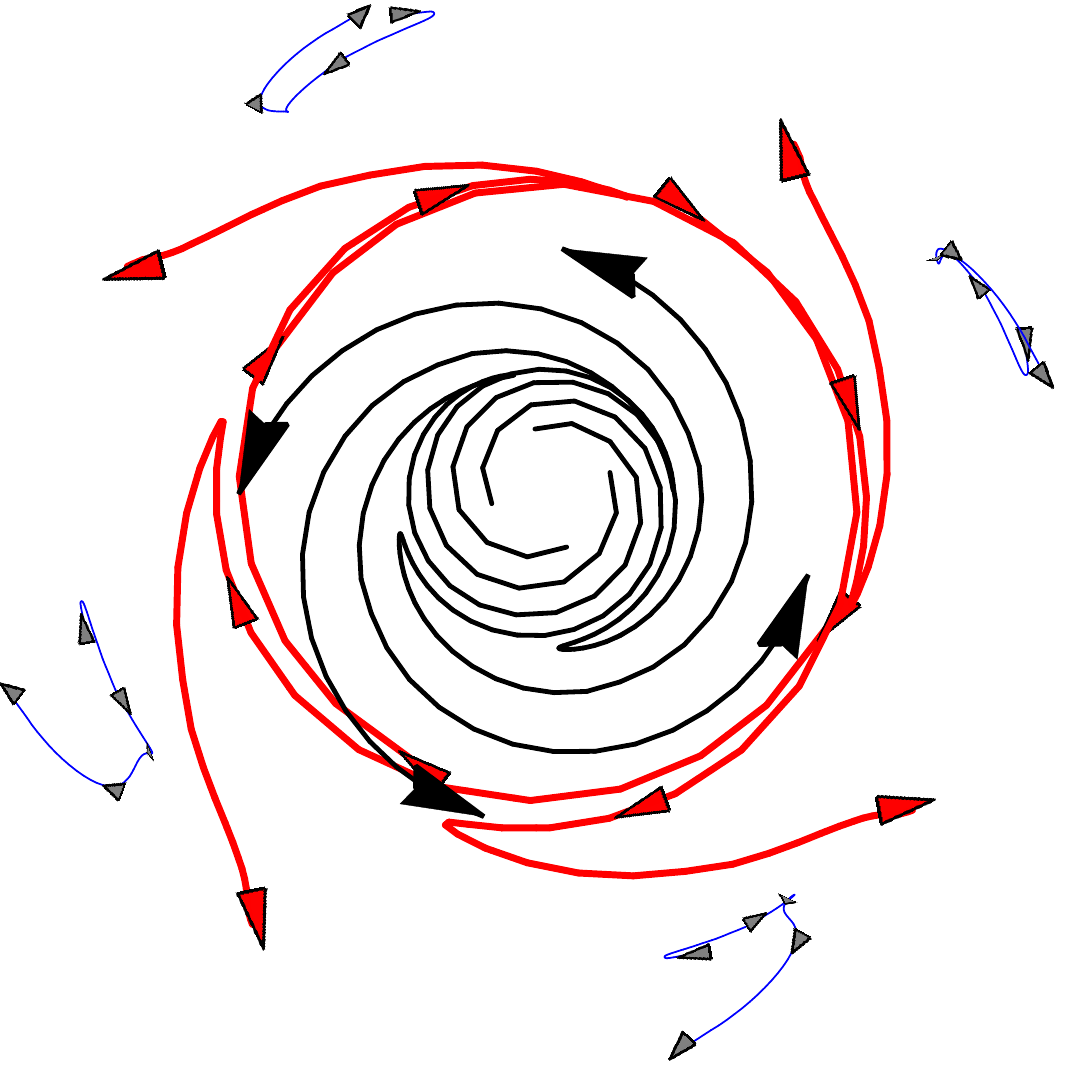}
        (b) top view
    \end{minipage}\hspace{0.5cm}
    \begin{minipage}{.3\linewidth}
        \centering
        \includegraphics[width=\textwidth]{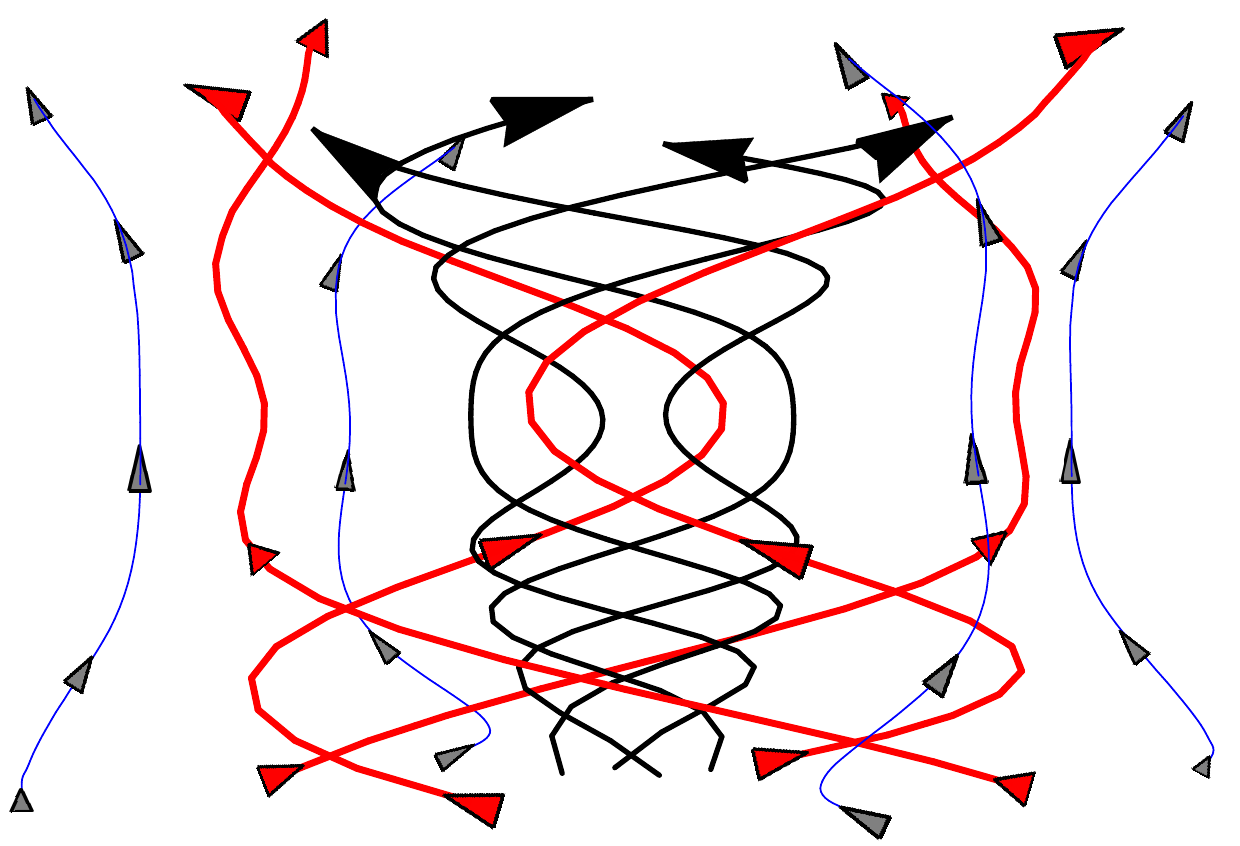}
        (c) front view
    \end{minipage}
\end{center}    
\end{minipage}    
\end{center}
    \caption[Stream lines.]{Vortex's stream lines.}
\label{streamLinesFig}
\end{figure}\\
It implies that $\phi$-velocity does not grow faster than $1/r$ as $r$ goes to zero, hence
\[
v_\phi(r,\phi,z) = \frac{1}{r}\int_{0}^r r'\omega_z(r',z)dr.
\label{vphiRadialIntegralOfOmegaEq}
\]
Since the $z$-vorticity does not have dependence with the $\phi$ coordinate, then the tangent velocity depends on $r$ and $z$. We may use Eq.~(\ref{vorticityPunctualChargeApproximationEq}) to find the following integral solution 
\begin{equation}
v_\phi(r,z) = -\frac{1}{r}\frac{\Omega}{\pi\nu}\lim_{N\to\infty} \sum_{\vec{r}_{\alpha} \in \mathcal{M}_N}  \sqrt{r_{\alpha}} \tilde{q}_\alpha \int_{0}^r \sqrt{r'} \mu(\pvec{r}',\vec{r}_{\alpha}) K[\mu(\pvec{r}',\vec{r}_{\alpha})] dr'.
\label{phiVelocityPunctualChargeApproximationEq}
\end{equation}
The computation of the tangent velocity using Eq.~(\ref{phiVelocityPunctualChargeApproximationEq}) requires the evaluation of the following integral 
\[
\mathcal{I}_\alpha (r)=\int_{0}^r \sqrt{r'} \mu(\pvec{r}',\vec{r}_{\alpha}) K[\mu(\pvec{r}',\vec{r}_{\alpha})] dr'
\]
whose exact solution is unknown. We have used a trapezoidal rule  \cite{davis2007methods} to evaluate $\mathcal{I}_\alpha (r)$ numerically. A plot of the tangent velocity at different values of $z>0$ according to Eq.~(\ref{phiVelocityPunctualChargeApproximationEq}) is shown in Fig.~\ref{rAndPhiVelocitiesFig}. The extreme values of the tangent velocity for $z/a = 0.4234, 0.3234 \mbox{ and } 0.2234$ are located at $z^{*}/a = 0.21 \pm 0.005, 0.22 \pm 0.005 \mbox{ and } 0.23 \pm 0.005$. Note that extreme values for the radial and tangent velocity are close to each other and they share a similar shifting property. We may use these extremes to define the average radius of the vortex core as a function of $z$. Since the tangent velocity is computed from a radial integration of the $z$-vorticity (see Eq.~(\ref{vphiRadialIntegralOfOmegaEq})), then $v_{\phi}(r, - z) = - v_{\phi}(r, - z)$ as occurs with the radial velocity. However, the sign of the radial velocity also depends on the sign of the angular velocity of the non-inertial reference frame. For example, $\Omega>0$ then we have the following behavior: then fluid enters and spins counterclockwise ($v_{\phi}(r,z)>0$) for $z<0$. On the other hand, the fluid exits and spins clockwise for $z>0$. The opposite situation occurs when $\Omega<0$, this is, a clockwise and counterclockwise spinning for $z<0$ and $z>0$ respectively. Vector field projections on several planes perpendicular to the $z$-axis are shown in Fig.~\ref{velocitySectionsFig} to illustrate this spinning behavior of the fluid as $z$-changes. There is no projection of the velocity field at the plane $z=0$ since $\omega_z$, $v_r$ and $v_\phi$ vanish and fluid flows vertically ($v_z(r,0)>0$). Finally, we have computed a few streamlines associated to the velocity field by using a standard Runge-Kutta scheme \cite{lambert1991numerical} with starting points located at the bottom of the vortex as it is shown in Fig.~\ref{streamLinesFig}.  

\section{Numerical solution of the Poisson's Equation}
\label{Numerical_solution_of_the_Poisson_EquationEqSection}
The purpose of the current section is to solve the Poisson's equation for the $z$-vorticity in order to do a comparison between the analytic approximation given by Eq.~(\ref{vorticityPunctualChargeApproximationEq}) and numerical solutions of Eq.~(\ref{steadyzVorticityEq}) via the \textit{Finite Difference Method} FDM \cite{chung2010computational, anderson1995computational}. We shall solve the Poisson's equation for $\omega_z$ in cylindrical coordinates 
\begin{equation}
\frac{1}{r}\partial_r \left( r \partial_r \omega_z \right) + \partial_z\partial_z \omega_z = \frac{2\Omega}{\nu}\Pi
\label{poissonEquationForZVorticityAxiallySymmEq}
\end{equation}
limiting the analysis for the axially symmetric problem $\omega_z=\omega_z(r,z)$ on a region \[
\tilde{D} = \left\{(r,z) : 0 < r < R \wedge |z| < H/2 \right\}\in\mathbb{R}^2
\]  
where $\Upsilon \subset \tilde{D}$ and $\partial\tilde{D} = \left\{(R,z) : |z| < H/2 \right\}\cup \left\{(r,H/2) : 0 \leq r \leq R \right\}\cup \left\{(r,-H/2) : 0 \leq r \leq R \right\}$ is the boundary. The corresponding finite difference approximation of Eq.~(\ref{poissonEquationForZVorticityAxiallySymmEq}) may be written as follows 
\[
\sum_{w \in \mathcal{D} } \mathcal{A}_{u w} (\omega_z)_w = \frac{2\Omega}{\nu}\Pi_{u}
\]
where $u=1,\ldots,N_r N_{z}$ label the nodes in a $N_r \times N_{z}$ rectangular lattice on the $rz$-plane whose nodes are located at $(r_i,z_j)=(i\delta r, -H/2 + j \delta z)$ and $\delta r \delta z$ is the area of the rectangular cell. The elements of the square matrix $\mathcal{A}_{u w}$ are given by 
\[
\mathcal{A}_{u w} =
   \begin{cases} 
      \sum_{\alpha=1}^2 \frac{1}{\delta \chi_{\alpha}}\left[\left(\sum_{\sigma\in \left\{+,-\right\} } \delta_{u_{(\alpha)}^\sigma,w}\right) - 2\delta_{u,w}\right] + \frac{1}{2\delta \chi_{1}(\chi_1)_{i(u)}}\sum_{\sigma\in \left\{+,-\right\} } \sigma \delta_{u_{(1)}^\sigma,w} & r_{u} \neq 0 \\
      \left(\frac{2}{\delta \chi_1}\right)^2\left(\delta_{u_{(1)}^{+},w}-\delta_{u,w}\right)+\left(\frac{1}{\delta \chi_2^2}\right)\left(\delta_{u_{(1)}^{+},w}+\delta_{u_{(1)}^{-},w}-2\delta_{u,w}\right) & r_{u} = 0 
   \end{cases}
\]
where $\delta_{u_{(\alpha)}^{\pm},w}$ are translations operators on the lattice along the $r$ and $z$ axes, $\chi_1 = r$, $\chi_2 = z$ and $i(u)$ is a indexing vector (see Appendix \ref{Finite_difference_implementationAppendix}). We have written a code in MATHEMATICA \cite{math} in order to compute the matrix $\mathcal{A}_{u w}$ and its inverse since $  (\omega_z)_u = \frac{2\Omega}{\nu} \sum_{w \in \mathcal{D} } \mathcal{A}_{u w}^{-1} \Pi_{w}$  for the inner points of the lattice. This strategy is a viable choice to solve the Poisson's equation, however there are other useful approaches e.g. iterative relaxation methods \cite{fox1948short,birkhoff1984numerical}, solutions obtained with the \textit{Boundary Element Method} (BEM) \cite{yildirim2008exact} or \textit{Finite Element Method} (FEM) solutions \cite{babuvska1973finite} especially for non-rectangular lattices.
\begin{figure}[h]
  \centering   
  \includegraphics[width=0.5\textwidth]{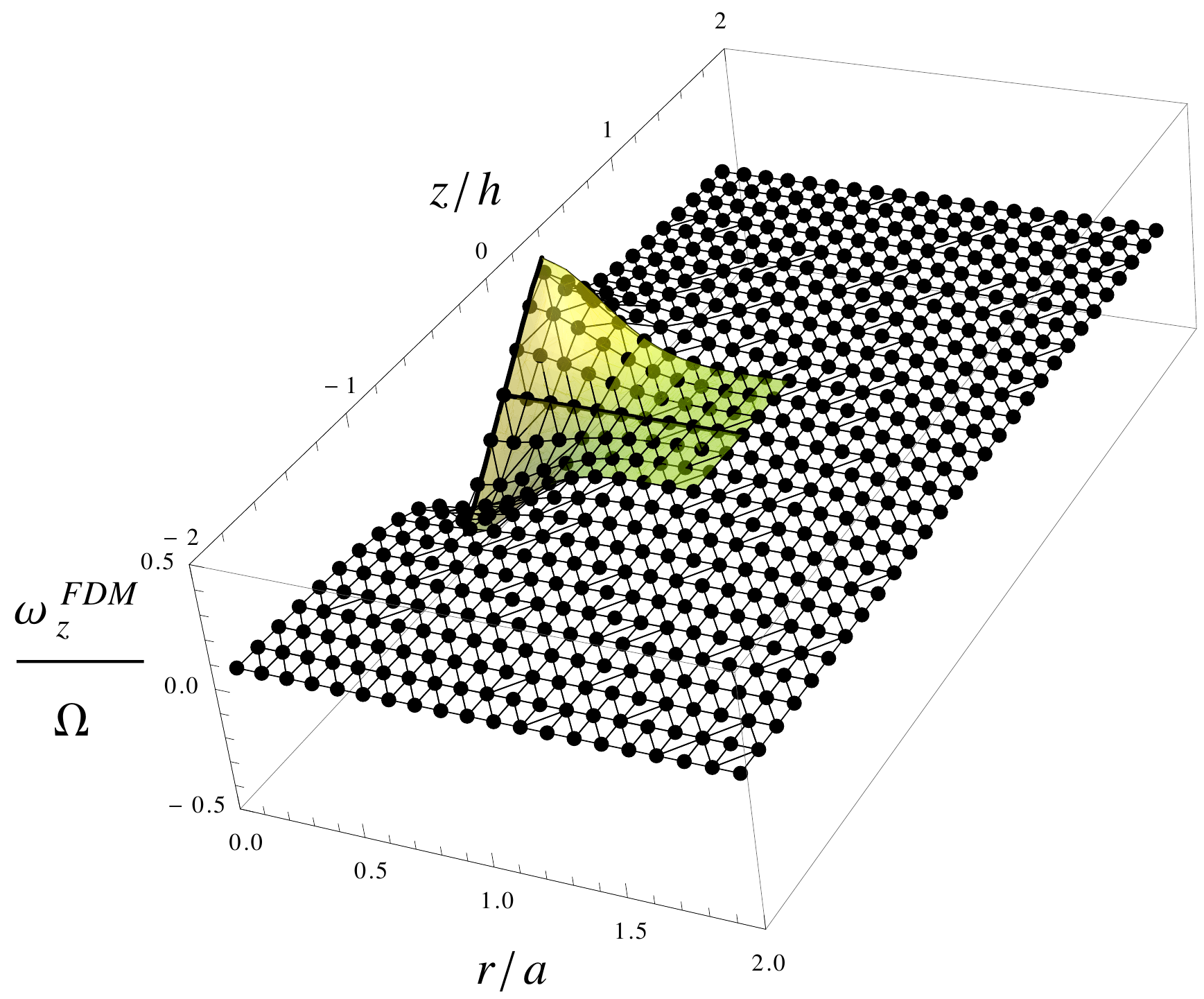}
  \includegraphics[width=0.45\textwidth]{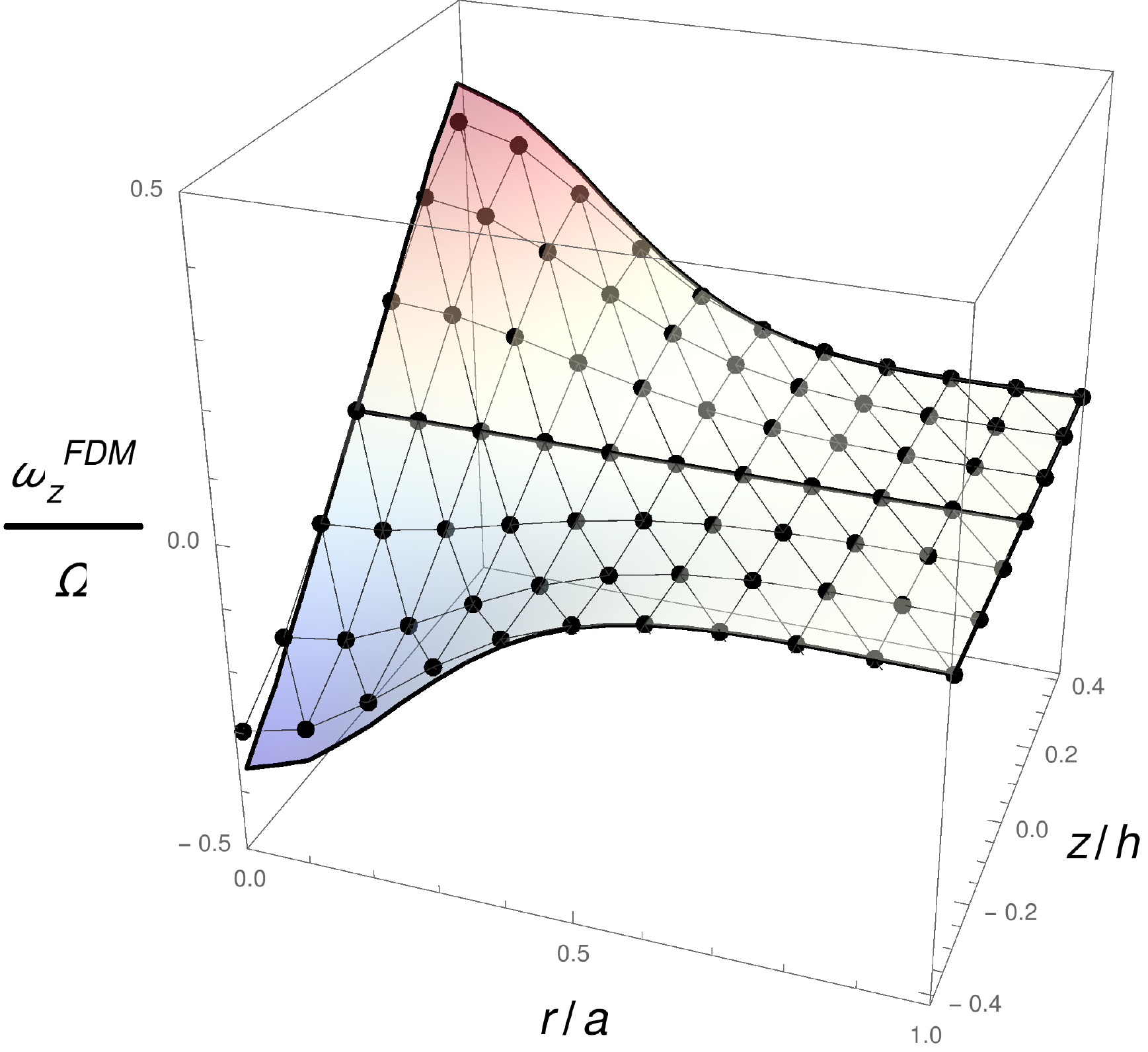}
  \caption[Comparison between numerical and the approximated analytical method.]%
  {\textbf{Comparison between numerical and the approximated analytical method.} In both plots the black points corresponds to the numerical FDM solution of the $z$-vorticity and the surface is represents the approximated analytic solution given by Eq.~(\ref{vorticityPunctualChargeApproximationEq}). \textbf{(left)} Solution in the region $\tilde{D}$. \textbf{(right)} Zoom of the solution in the region of interest $\Upsilon \subset \tilde{D}$.}
  \label{VorticityComparisonFig}
\end{figure}\\
A comparison between the numerical and approximated analytical method for the $z$-vorticity is shown in Fig.~\ref{VorticityComparisonFig}. The FDM computation was performed with a $20 \times 30$ lattice setting $H/h=4$ and $R/a=2$. The numerical and approximated analytical results are in good agreement except the near the global extreme of $\omega_z$ since the FDM tries to adapt the solution to the Dirichlet boundary condition of zero vorticity along $\partial\tilde{D}$. This deviation occurs because the FDM solution is obtained in a finite lattice while the approximated solution of Eq.~(\ref{poissonEquationForZVorticityAxiallySymmEq}) was found by moving the boundaries to the infinity. Both solutions will have the same maximum values as $H/h>>1$ and $R/a>>1$, but this demands larger FDM computation to keep simulations with a decent density of points on the lattice.           

\section{Concluding remarks}
In this document, we have presented an analytic approximation for rotating Stokes flows. The approach may be used when the vorticity is highly concentrated along the axis of rotation. In the framework of this approximation, any $z$-velocity field satisfying the Laplace equation may be used to determine the vorticity. Integral solutions for the vorticity were found in terms of the elliptic functions of the first and second kind. These expressions were evaluated and compared with numerical solutions obtained with FDM. Although, the approximated $z$-vorticity given by Eq.~(\ref{vorticityPunctualChargeApproximationEq}) and the FDM solutions are in good agreement for the axially symmetric case, we found that the maximum value of the numerical vorticity $|\omega_z^{FDM}|$ was slightly lower than the one found with Eq.~(\ref{vorticityPunctualChargeApproximationEq}) suggesting that boundaries of the rectangular lattice (see Fig.~\ref{VorticityComparisonFig}-\textbf{(left)}) should be move to the infinite to avoid that Dirichlet boundary conditions of zero vorticity change drastically such extreme values. 

Even when we deal with a highly viscous problem, this study provides insights about the structural properties of three-dimensional vortices in the atmosphere as the vorticity sign inversion commonly occurring on the cross-section of tropical cyclones along their axis of rotation.

\section*{Acknowledgments}
This work was supported by the Vicerrector\'ia de investigaciones, Universidad ECCI and the Direcci\'on de ciencias b\'asicas, Universidad ECCI.

\begin{appendices}
\section{Finite difference method implementation}
\label{Finite_difference_implementationAppendix}
Assume that we have a lattice whose nodes are located at a rectangular lattice in the $rz$-plane. If $N_r$ and $N_z$ are the points along the $r$ and $z$ directions, then the node positions may be defined as $(r_i,z_j)=(i\delta r, -H/2 + j \delta z)$ with $\delta r = R/N_r$ and $\delta z = H/N_z$ respectively. The central finite difference scheme for $r \neq 0$ is given by
\[
\left. \partial_r \omega_z \right|_{(i,j)} = \frac{1}{2 r_i \delta r}\left[ (\omega_z)_{i+1,j} - (\omega_z)_{i-1,j} \right] + O(\delta r)\hspace{1.0cm}\mbox{,}  
\] 

\[
\left. \partial_r\partial_r \omega_z \right|_{(i,j)} = \frac{1}{\delta r^2}\left[ (\omega_z)_{i+1,j} + (\omega_z)_{i-1,j} - 2(\omega_z)_{i,j}\right] + O(\delta r^2)  
\]
and
\[
\left. \partial_z\partial_z \omega_z \right|_{(i,j)} = \frac{1}{\delta z^2}\left[ (\omega_z)_{i,j+1} + (\omega_z)_{i,j-1} - 2(\omega_z)_{i,j}\right] + O(\delta z^2).  
\]
We use an indexing function $u=u(i,j)=1,\ldots,\mathcal{N}$ with $\mathcal{N}=N_r N_z$ to write the central derivatives in terms of a single index. Additionally, it is convenient to define the following operator
\begin{equation}
u_{(\alpha)}^{\pm} = u\left( (i(u),j(u)) \pm e_\alpha \right) 
\label{latticeOperatorEq}
\end{equation}
with
\[
e_1 := (1,0)\hspace{0.5cm}\mbox{and}\hspace{0.5cm}e_2 := (0,1)
\]
the unit vectors on the $rz$-plane. The operator defined in Eq.~(\ref{latticeOperatorEq}) represents a movement from the mesh node $(i,j)$ to one of its nearest-neighbouring nodes. Then the Laplacian operator on an axially symmetric function may be written as follows    
\[
\left.\partial_{\alpha}\partial_{\alpha}(\omega_z(r,z))\right|_{(i,j)} = \sum_{w \in \mathcal{D} } a_{u w} (\omega_z)_w \hspace{0.5cm} \mbox{\textbf{if}} \hspace{0.5cm} r_{i(u)} \neq 0
\]
where we have defined
\[
a_{u w} = \sum_{\alpha=1}^2 \frac{1}{\delta \chi_{\alpha}}\left[\left(\sum_{\sigma\in \left\{+,-\right\} } \delta_{u_{(\alpha)}^\sigma,w}\right) - 2\delta_{u,w}\right] + \frac{1}{2\delta \chi_{1}(\chi_1)_{i(u)}}\sum_{\sigma\in \left\{+,-\right\} } \sigma \delta_{u_{(1)}^\sigma,w}
\]
and $\chi_1 = r$, $\chi_2 = z$. We stress the fact that cylindrical coordinates require a special treatment at the origin because nodes at $r=0$ are not boundary points. Since inner points nodes on the $r$-axis are unknowns, then they must satisfy the Poisson's equation. Let us consider a tiny cylindrical region $V(\epsilon,\delta) := \left\{ (r,\phi,z) : r < \epsilon, |z| < \delta/2 \mbox{ and } 0 < \phi \leq 2\pi \right\}$ centred at $(0,z)$, then  
\[
\int_{V(\epsilon,\delta)}\left[ \frac{1}{r}\partial_r \left( r \partial_r \omega_z \right) + \partial_z\partial_z \omega_z(r,z)\right] dV = \frac{2\Omega}{\nu} \int_{V(\epsilon,\delta)} \Pi(r,z) dV
\]
Now
\[
\int_{V(\epsilon,\delta)}\left[ \frac{1}{r}\partial_r \left( r \partial_r \omega_z \right) \right] dV = \int_{0}^{2\pi} d\phi \int_{-\delta/2}^{\delta/2}dz \int_{0}^\epsilon r dr \left[ \frac{1}{r}\partial_r \left( r \partial_r \omega_z \right) \right] = 2\pi \int_{-\delta/2}^{\delta/2} dz \left[ \epsilon \partial_r \omega_z(\epsilon,z) \right]  
\]
Here is convenient to take $\epsilon=\delta r/2$ and $\delta = 2 \delta z$ in order to apply central differences in the radial to obtain
\[
\int_{V(\epsilon,\delta)}\left[ \frac{1}{r}\partial_r \left( r \partial_r \omega_z \right) \right] dV \underset{V(\epsilon,\delta) \rightarrow 0}{=} 2\pi (2 \delta z) \left[ \left(\frac{\delta r}{2}\right) \frac{\omega_z(\delta r,z)-\omega_z(0,z)}{\delta r} \right] .  
\]
On the other hand, the integral over the term $\partial_z\partial_z \omega_z(r,z)$ can be approximate as follows
\[
\int_{V(\epsilon,\delta)}\left[ \partial_z\partial_z \omega_z(r,z)\right] dV \underset{V(\epsilon,\delta) \rightarrow 0}{=}  \delta V \left[ \partial_z\partial_z \omega_z(0,z)\right] = \left[2\pi \frac{\delta r}{2} (2 \delta z) \right]\left[ \frac{\omega_z(0,z+\delta z) + \omega_z(0,z+\delta z) - 2\omega_z(0,z) }{\delta z^2}\right]
\]
where we have applied central difference on the term $\partial_z\partial_z \omega_z(0,z)$. Similarly, the integral over the sink/source vorticity term can be calculated as follows 
\[
\int_{V(\epsilon,\delta)} \Pi(r,z) dV \underset{V(\epsilon,\delta) \rightarrow 0}{=} \delta V \Pi(0,z) = \left[2\pi (2 \delta z) \left(\frac{\delta r}{2}\right)\right] \Pi(0,z).
\]
Therefore, the Poisson's equation takes the form
\[
\left(\frac{2}{\delta r}\right)^2 \left[(\omega_z)_{u(1,j)} - (\omega_z)_{u(0,j)}\right] + \left(\frac{1}{\delta z^2}\right) \left[(\omega_z)_{u(0,j+1)} + (\omega_z)_{u(0,j-1)} - 2(\omega_z)_{u(0,j)}\right]=\frac{2\Omega}{\nu}\Pi_{u(0,j)}.
\]
The previous expression may be written as 
\[
\left(\frac{2}{\delta r}\right)^2 \left[(\omega_z)_{u_{(1)}^+} - (\omega_z)_{u}\right] + \left(\frac{1}{\delta z^2}\right) \left[(\omega_z)_{u_{(2)}^+} + (\omega_z)_{u_{(2)}^-} - 2(\omega_z)_{u}\right]=\frac{2\Omega}{\nu}\Pi_{u}\hspace{0.5cm} \mbox{\textbf{if}} \hspace{0.5cm}r_{i(u)} = 0
\]
by using Eq.~(\ref{latticeOperatorEq}) hence
\[
\left.\partial_{\alpha}\partial_{\alpha}(\omega_z(r,z))\right|_{(i,j)} = \sum_{w \in \mathcal{D} } b_{u w} (\omega_z)_w \hspace{0.5cm} \mbox{\textbf{if}} \hspace{0.5cm} r_{i(u)} = 0
\]
where
\[
b_{u w} = \left(\frac{2}{\delta \chi_1}\right)^2\left(\delta_{u_{(1)}^{+},w}-\delta_{u,w}\right)+\left(\frac{1}{\delta \chi_2^2}\right)\left(\delta_{u_{(1)}^{+},w}+\delta_{u_{(1)}^{-},w}-2\delta_{u,w}\right) .
\]
Finally,
\[
\left.\partial_{\alpha}\partial_{\alpha}(\omega_z(r,z))\right|_{u} = \frac{2\Omega}{\nu}\Pi_{u}
\]
can be written as
\[
\sum_{w \in \mathcal{D} } \mathcal{A}_{u w} (\omega_z)_w = \frac{2\Omega}{\nu}\Pi_{u}
\]
where
\[
\mathcal{A}_{u w} = a_{u w} \hspace{0.5cm} \mbox{\textbf{if}} \hspace{0.5cm} (\chi_1)_{i(u)} \neq 0 \hspace{0.5cm} \mbox{\textbf{otherwise}} \hspace{0.5cm} b_{u w} .
\]
Note that definition of Eq.~(\ref{latticeOperatorEq}) allows to write straightforwardly linear operators as the Laplacian as a matrix on the lattice and the Poisson's equation can be written as a linear set of equations. The technique described here is also useful in other two-dimensional problems including the Laplacian as the Helmholtz equation \cite{salazar2015chaos} where the problem is reduced to find the eigenvalues of a matrix with entries defined on the two-dimensional lattice.

\end{appendices}

\bibliographystyle{ieeetr} 
\bibliography{bibliography.bib}






\end{document}